\documentclass[nofootinbib,a4paper,aps,prd,10pt,superscriptaddress,showkeys,twocolumn]{revtex4}
\usepackage{graphicx}
\usepackage{amsfonts}
\usepackage{amssymb} 
\usepackage{amsmath}
\usepackage{hyperref}
\usepackage{natbib}
\usepackage{float}
\usepackage{dcolumn}
\usepackage{bm}
\usepackage{latexsym,color}

\def\mnras{Mon. Not. Roy. Astr. Soc.}
\def\apj{Astrophys. J.}
\def\apjl{Astrophys. J. Lett.}

\def\aap{Astron. Astrophys.}

\def\araa{Annual Rev. of Astron. Astrophys.}

\def\nphysa{Nucl. Phys.}
\def\physrep{Phys. Rep.}

\begin{document}

\title{Accretion disk in the Hartle-Thorne spacetime}

\author{Yergali~\surname{Kurmanov}}
\email[]{kurmanov.yergali@kaznu.kz}
\affiliation{National Nanotechnology Laboratory of Open Type,  Almaty 050040, Kazakhstan.}
\affiliation{Al-Farabi Kazakh National University, Al-Farabi av. 71, 050040 Almaty, Kazakhstan.}

\author{Marco Muccino}
\email[]{marco.muccino@lnf.infn.it}
\affiliation{Universit\`a di Camerino, Via Madonna delle Carceri 9, 62032 Camerino, Italy.}
\affiliation{Al-Farabi Kazakh National University, Al-Farabi av. 71, 050040 Almaty, Kazakhstan.}

\author{Kuantay~\surname{Boshkayev}}
\email[]{kuantay@mail.ru}
\affiliation{National Nanotechnology Laboratory of Open Type,  Almaty 050040, Kazakhstan.}
\affiliation{Al-Farabi Kazakh National University, Al-Farabi av. 71, 050040 Almaty, Kazakhstan.}
\affiliation{International Information Technology University, Manas st. 34/1, 050040 Almaty, Kazakhstan.}

\author{Talgar~\surname{Konysbayev}}
\email[] {talgar\_777@mail.ru}
\affiliation{National Nanotechnology Laboratory of Open Type,  Almaty 050040, Kazakhstan.}
\affiliation{Al-Farabi Kazakh National University, Al-Farabi av. 71, 050040 Almaty, Kazakhstan.}

\author{Orlando~\surname{Luongo}}
\email[]{orlando.luongo@unicam.it}
\affiliation{Al-Farabi Kazakh National University, Al-Farabi av. 71, 050040 Almaty, Kazakhstan.}
\affiliation{Universit\`a di Camerino, Via Madonna delle Carceri 9, 62032 Camerino, Italy.}
\affiliation{SUNY Polytechnic Institute, 13502 Utica, New York, USA.}
\affiliation{Istituto Nazionale di Fisica Nucleare, Sezione di Perugia, 06123, Perugia,  Italy.}
\affiliation{INAF - Osservatorio Astronomico di Brera, Milano, Italy.}

\author{Hernando~\surname{Quevedo}}
\email[]{quevedo@nucleares.unam.mx}
\affiliation{Al-Farabi Kazakh National University, Al-Farabi av. 71, 050040 Almaty, Kazakhstan.}
\affiliation{Instituto de Ciencias Nucleares, Universidad Nacional Aut\`onoma de M\`exico, Mexico. }
\affiliation{Dipartimento di Fisica and ICRA, Universit\`a di Roma “La Sapienza”, Roma, Italy.}

\author{Ainur~\surname{Urazalina}}
\email[]{y.a.a.707@mail.ru}
\affiliation{National Nanotechnology Laboratory of Open Type,  Almaty 050040, Kazakhstan.}
\affiliation{Al-Farabi Kazakh National University, Al-Farabi av. 71, 050040 Almaty, Kazakhstan.}

\begin{abstract}
We consider the circular motion of test particles in the gravitational field  of a rotating deformed object described by the Hartle-Thorne metric. This metric represents an approximate solution to the vacuum Einstein field equations, accurate to second order in the angular momentum $J$ and to first order in the mass quadrupole moment $Q$. We calculate the orbital parameters of neutral test particles on circular orbits (in accretion disks) such as angular velocity, $\Omega$, total energy, $E$, angular momentum, $L$, and radius of the innermost stable circular orbit, $R_{ISCO}$, as functions of the total mass, $M$, spin parameter, $j=J/M^2$ and quadrupole  parameter, $q=Q/M^3$, of the source. We use the Novikov-Thorne-Page thin accretion disk model to investigate  the characteristics of the disk. In particular, we analyze in detail the radiative flux, differential luminosity, and spectral luminosity of the accretion disk, which are the quantities that can be  measured experimentally. We compare our results with those obtained in the literature for the Schwarzschild and Kerr metrics, and the $q$-metric. It turns out that the Hartle-Thorne metric and the Kerr metric lead to similar results for the predicted flux and the differential and spectral luminosities, whereas the q-metric predicts different values. We compare the predicted values of $M$, $j$, and $q$ with those of realistic neutron star models. Furthermore, we compare the values of $R_{ISCO}$ with the static and rotating radii of neutron stars. 
\end{abstract}


\maketitle
 
\section{Introduction}

The influx of gas and dust or, more general, diffuse material, towards a central gravitating object is dubbed \emph{accretion}, mostly occurring through the formation of \emph{accretion disks} \cite{1971reas.book.....Z,1973grav.book.....M, 1983bhwd.book.....S}. The formation of an accretion disk is unquestionably one of the most prevalent processes in relativistic astrophysics and, importantly, it yields significant observational manifestations. Notably, the accretion of matter onto relativistic objects, such as black holes and neutron stars, stands out as one of the most efficient mechanisms for energy release in the field of astrophysics \cite{2013grsp.book.....O}. Undoubtedly, the most remarkable observational manifestation is the accretion onto a black hole that results in the release of an enormous amount of energy per unit of accreted mass and showcases the dynamics associated with black hole accretion.

Accretion disks enable the observation of the radiation emitted by matter in rotational motion around a compact object. By analyzing the emitted spectra of the disk, valuable information about the nature of the central object undergoing accretion can be obtained. These observations provide insights into the properties and characteristics of the accreting central object, contributing to our understanding of astrophysical phenomena.

The exploration of central compact objects, which also includes supermassive central objects, carries immense importance in contemporary astrophysics. The origin of these objects is still a topic of ongoing discussion. Comprehending their physical attributes necessitates indirect research methods, such as examining \emph{accretion disks}. Accretion disks play a crucial role in uncovering the fundamental  properties of the aforementioned structures. By conducting thorough analyses and observations of accretion disks, we can acquire valuable knowledge about the development and dynamics of central compact objects\footnote{Nevertheless, the incoming observations can also shed some new light on possible deviations from general relativity in the incoming years. Consequently, the possible existence of exotic compact objects cannot be ignored, as most observations of black hole candidates do not allow one to study the geometry near such astrophysical sources yet.}. 

Accretion disk luminosity for compact objects can be modeled using various solutions to Einstein's field equations. This includes the gravitational field of neutral black holes, as described in Ref. \cite{2012LRR....15....7C}, or the outside field of white dwarfs and neutron stars, as discussed in Refs. \cite{1983bhwd.book.....S,haenselbook,2013ApJ...762..117B, 2014NuPhA.921...33B}. Additionally, exotic objects such as boson stars \cite{2002NuPhB.626..377T,2006PhRvD..73b1501G,beheshagasp,cardopan} or gravastars \cite{2004PNAS..101.9545M} can also be considered. Observations related to accretion disks provide valuable insights into these objects as well. Examples include the motion of stars near the galactic center \cite{2018A&A...615L..15G, 1998ApJ...509..678G, 2000Natur.407..349G}, the spectra of X-ray binary systems \cite{x-ray} and the emission by the accretion disks of binary black holes \cite{2016PhRvL.116f1102A,2016PhRvL.116v1101A}. Other examples encompass the shadow of supermassive black hole candidates in the nucleus of the M87 galaxy \cite{2019ApJ...875L...1E}, among others. It is worth noting that certain features of the accretion disk depend on the underlying space-time geometry \cite{2016gac..conf..185A}. Hence, observations of accretion disks can be utilized to impose constraints on the geometry, as discussed in \cite{abramowicz}. By comparing theoretical models with observational data, we can gain insights into the nature of compact objects and the properties of the space-time in their vicinity.

Among all possible spacetimes modelling compact objects, from which we can infer accretion disk properties, the Hartle-Thorne metric represents a useful tool for describing the geometry around slowly rotating and slightly deformed objects in strong gravitational fields. As stated, it provides a framework for studying real astrophysical objects ranging from celestial bodies like planets to neutron stars. The metric is characterized by three multipole moments: the total mass, the angular and quadrupole momenta. These parameters describe several astrophysical phenomena \cite{2001CQGra..18..969A,2003LRR.....6....3S,2004MNRAS.350.1416B,2005MNRAS.358..923B}. 

One notable advantage of the Hartle-Thorne metric is its flexibility. In cases where angular momentum is absent, it characterizes a naked singularity. However, by incorporating angular momentum and expressing the quadrupole moment based on it, the metric transforms into the well-known Kerr metric. When linear terms in angular momentum are present without the quadrupole moment, it corresponds to the Lense-Thirring solution. Additionally, in the absence of angular momentum, a comparison can be made between the Hartle-Thorne metric and the approximate Erez-Rosen solution \cite{1991NCimB.106..545M}. Consequently, the Hartle-Thorne solution serves as a valuable reference point for modeling intricate variations of central compact objects.

In this work, we construct a theoretical scheme to determine the accretion disk, following the standard theory of black hole accretion developed in \cite{novikov1973, page1974}, which  can be adapted to the Hartle-Thorne solution within the context of  Einstein's equations. So, we first compute the particle motion in such a metric, emphasizing the role of the aforementioned  three free parameters. Afterwards, we determine the circular orbits, the innermost stable circular orbits (ISCOs) and the kinematic properties of the metric, used to infer the spectral properties of thin accretion disks. We compare our findings with those of the Schwarzschild and Kerr metrics as limiting cases of the Hartle-Thorne spacetime. We motivate this, noticing that specific properties of given compact objects can modify the form of the metric and so,  utilizing the Hartle-Thorne metric, one can compute the luminosity and fluxes associated with these central compact objects. These calculations provide valuable insights into the observable properties of accretion processes and the radiative emissions from the surrounding matter. Finally, theoretical interpretations of our outcomes are critically discussed, employing  generic neutron star models that are compared with our framework.  

The paper is organized as follows. In Sect. \ref{sez2}, we review the Hartle-Thorne metric and  compare it with other known solutions, In Sect. \ref{sez3}, we consider the circular orbits and basic parameters of neutral test particles in the Hartle-Thorne spacetime. In Sect. \ref{sez4},  we review the Novikov-Page-Thorne model and present our numerical results for the spectral properties of the accretion disks. In Sect.~\ref{sec:ns}, we discuss about neutron star physics. Finally, in Sect. \ref{sez6}, we present the  conclusions and perspectives of our work. Throughout the paper we make use of geometrized units setting $G=c=1$.


\section{Particle motion in the Hartle-Thorne metric}\label{sez2}

The line element for the Hartle-Thorne metric is \cite{2009CQGra..26v5006B} \\

\begin{widetext}
\begin{align}\label{ht1}
ds^2=\,&-\left(1-\frac{2{ M }}{r}\right) \left[1+2k_1P_2(\cos\theta)+2\left(1-\frac{2{M}}{r}\right)^{-1} \frac{J^{2}}{r^{4}}(2\cos^2\theta-1)\right]dt^2+\left(1-\frac{2{M}}{r}\right)^{-1}\nonumber\\
&\times\left[1-2k_2P_2(\cos\theta)-2\left(1-\frac{2{M}}{r}\right)^{-1}\frac{J^{2}}{r^4}\right]dr^2
+r^2[1-2k_3P_2(\cos\theta)](d\theta^2+\sin^2\theta d\phi^2) -\frac{4J}{r}\sin^2\theta dt d\phi\,,
\end{align}
\end{widetext}
\noindent with
\begin{eqnarray}\label{k123}
k_1&=&\frac{J^{2}}{{M}r^3}\left(1+\frac{{M}}{r}\right)+\frac{5}{8}\frac{Q-J^{2}/{M}}{{M}^3}Q_2^2\left(\frac{r}{{M}}-1\right),\nonumber\\
k_2&=&k_1-\frac{6J^{2}}{r^4},\nonumber\\
k_3&=&k_1+\frac{J^{2}}{r^4}+\frac{5}{4}\frac{Q-J^{2}/{M}}{{M}^2\left(r^2-2Mr\right)^{1/2}} Q_2^1\left(\frac{r}{M}-1\right)\,,\nonumber
\end{eqnarray}
and
\begin{eqnarray}\label{ht2}
P_{2}(\cos\theta)&=&\frac{1}{2}(3\cos^{2}\theta-1),\nonumber\\
Q_{2}^{1}(x)&=&(x^{2}-1)^{1/2}\left[\frac{3x}{2}\ln\frac{x+1}{x-1}-\frac{3x^{2}-2}{x^{2}-1}\right],\nonumber\\
Q_{2}^{2}(x)&=&(x^{2}-1)\left[\frac{3}{2}\ln\frac{x+1}{x-1}-\frac{3x^{3}-5x}{(x^{2}-1)^2}\right],\nonumber
\end{eqnarray}
where $x=r/M-1$, $P_{2}(x)$ is the second Legendre polynomial of the first kind, $Q_l^m$ are the associated Legendre polynomials of the second kind and the constants ${M}$, ${J}$ and ${Q}$ are, as mentioned earlier, the total mass, angular momentum and mass quadrupole moment of a rotating star, respectively. In addition, $J\sim\Omega_{Star}$ and $Q\sim\Omega_{Star}^2$, where $\Omega_{Star}$ is the angular velocity of the central object, or a star, behaving as the source of gravity.

In general, the Hartle-Thorne metric encompasses both interior and exterior solutions that describe the gravitational field within and outside a compact object, respectively. However, when one investigates the accretion disk, focusing on the exterior case appears sufficient. Indeed, this occurs because the accretion disk is primarily influenced by the gravitational field outside the compact object. However, there are certain situations where it becomes necessary to explore the interior gravitational field of a compact object as well as the exterior one. This is particularly relevant when constructing quantities such as mass-radius relations or mass-central density profiles. In such cases, it is important to appropriately consider both the interior and exterior solutions of the Hartle-Thorne metric in order to obtain a complete description of the compact object.


\subsection{Comparison with alternative spacetimes} \label{sec:comparison}

Here we review the relationship of the Hartle-Thorne metric with other solutions in the literature, survey studies on accretion disks in those spacetimes and highlight some open issues in this direction.

The exterior Hartle-Thorne solution characterizes the geometry surrounding compact objects with slow rotation and slight deformations. In their groundbreaking paper \cite{1968ApJ...153..807H}, Hartle and Thorne initially attempted to compare this solution with the well-known Kerr solution. It has been demonstrated that by selecting a specific value for the quadrupole moment, denoted as $q=j^2$, and applying intricate coordinate transformations, the Hartle-Thorne solution converges to the approximate Kerr solution in the case of slow rotation. This implies that the applicability of the Kerr solution is limited and it can only describe a particular category of objects, namely the geometry around rotating black holes\footnote{The luminosity of the accretion disk in the Kerr spacetime has been extensively investigated in the scientific literature.}.

Subsequently,  the Hartle-Thorne metric can be compared with a static spacetime describing the geometry around deformed objects, in particular, with the Erez-Rosen metric\footnote{It was shown that the static Hartle-Thorne solution with $j=0$ reduces to the approximate Erez-Rosen solution in the limiting case of a small deformation \cite{1991NCimB.106..545M}. However, before finding the coordinate transformations to establish the relationship between the parameters of the solutions, it was necessary to generalize the Erez-Rosen metric by applying a Zipoy-Voorhees transformation, which introduces a new parameter that must be fixed in order to obtain the required transformations  \cite{1991NCimB.106..545M}.} \cite{2003esef.book.....S}.
Then, by using the Geroch-Hansen invariant definition of multipole moments \cite{1970JMP....11.2580G,1974JMP....15...46H} it was established that the monopole moment corresponds to the total mass, the dipole moment is the angular momentum, and the quadrupole moment contains an intrinsic part, due to the deformation of the source, and a second term due to  rotation.
As a result, it was shown that the Hartle-Thorne parameters $M$ and $Q$ are related to those of the Erez-Rosen metric as follows $M=M_{ER}(1-q_{ER})$ and $Q=-(4/5)q_{ER}M_{ER}^3$. This result allows one to find unambiguously the coordinate transformations between the static Hartle-Thorne and the approximate Erez-Rosen solutions\footnote{Some details about the Zipoy-Voorhees transformations that are necessary to compare the Erez-Rosen and Hartle-Thorne solutions can be found in Ref.~\cite{2019Symm...11.1324B}.}. 

A further attempt to find the relationship between the Hartle-Thorne and Erez-Rosen solutions was made by Frutos-Alfaro and Soffel in the limit of $\sim M^2$ and $\sim Q^2$ \cite{FrutosAlfaro2015OnTP}. To this end, the original exterior Hartle-Thorne solution was generalized to include new terms $\sim Q^2$ for a non-rotating case. The advantage of this approach is that it is not necessary to perform a Zipoy-Voorhees transformation in order  to obtain the corresponding coordinate transformations, which turned out to have a completely different form. It was concluded that the approximation determines the coordinate transformation (for more details see e.g. Ref.~\cite{boshkayev2020corr}). To the knowledge of the authors, the luminosity of accretion disks in the Erez-Rosen spacetime has not been studied yet in the literature.

Additionally, in Refs. \cite{1985PhLA..109...13Q,1989PhRvD..39.2904Q,1990ForPh..38..733Q, 1990PhLA..148..149Q,1991PhRvD..43.3902Q}, it was found a whole new class of exterior exact solutions with an infinite number of parameters, containing not only the mass, rotation parameter, and quadrupole parameter, but also the  Zipoy-Voorhees parameter, charge, and the Taub-NUT parameter, among others\footnote{This solution  contains as a specific case the solution combining both Erez Rosen and Kerr solutions. Using the prescriptions provided in Refs.~\cite{1968ApJ...153..807H,1991NCimB.106..545M},   it was demonstrated that this particular Quevedo-Mashhoon solution in the limiting case of slow rotation and small deformation was equivalent to the Hartle-Thorne solution \cite{2009CQGra..26v5006B}.}. In this respect, one of the main drawbacks of the exterior exact solutions is the fact that it is hard or sometimes even impossible to find their interior counterparts. Nonetheless, as above stated, for most of the astrophysical phenomena the exterior exact solutions are more than sufficient and play a pivotal and occasionally ultimate role. The luminosity of the accretion disk in the spacetime combining both Kerr and Erez-Rosen solutions has not been studied yet.

In the weak field regime, namely in the post-Newtonian approximation, it was shown that the Hartle-Thorne solution reduces to the Fock solution with quadrupole moment \cite{2012PhRvD..86f4043B}. The generalization of the Hartle-Thorne solution to quartic order in angular velocity  has been obtained in Ref.~\cite{2014PhRvD..89l4013Y}. Undoubtedly, the extension of the Hartle-Thorne solution may have a wider application in astrophysics\footnote{Additionally, it was shown that the so-called Sedrakyan-Chubaryan solution is equivalent to the the Hartle-Thorne solution \cite{2016GrCo...22..305B}.}.

In view of recent developments \cite{2019PhRvD..99d4005A}, it should also be possible to establish the relationship between the Hartle-Thorne solution and q-metric\footnote{Sometimes, this metric is known in the literature as the Zipoy-Voorhees metric, $\delta$-metric and $\gamma$-metric.} and its extension that includes the rotation parameter \cite{2018RSOS....580640F,2020CQGra..37e5006A}. However, this issue is out of the scope of the present work and possibly will be addressed in future studies.

Given the remarkable properties of the Hartle-Thorne spacetime and its widespread applications, it is highly intriguing to explore its implications in the context of accretion disks. By incorporating quadrupole terms and considering modifications to the standard Zipoy-Voorhees metric, we can account for additional factors such as mass and other properties that may influence the dynamics of the system. This allows us to extend the applicability of the Hartle-Thorne solution beyond static spherically symmetric configurations and incorporate rotation effects. Remarkably, the predictions derived from our analysis are anticipated to be tested experimentally in the near future, further validating the significance of our findings. In this respect, below we elucidate the main features of the Hartle-Thorne solution in view of observable signatures that can be obtained from it. 


\section{Circular orbits in the Hartle-Thorne space-time}\label{sez3}

We focus our analysis on the equatorial plane, where the polar angle $\theta$ is fixed at $\pi/2$. Within this restricted region, we investigate the motion of neutral test particles that follow circular orbits. By employing the well-established Euler-Lagrange formalism, we are able to determine various orbital parameters for these particles, including the angular velocity, angular momentum, and energy\footnote{The formulas for the angular velocity $\Omega$, angular momentum $L$, and energy $E$ in the Hartle-Thorne spacetime were initially derived in Ref.~\cite{2003gr.qc....12070A}. However, it is important to exercise caution when referring to \cite{2003gr.qc....12070A} due to the presence of certain typographical errors.}. These quantities play a crucial role in characterizing the dynamics of the test particles and provide valuable insights into their behavior within the considered gravitational field.

\subsection{Angular momentum and energy}

The angular velocity for corotating/counterrotating particles on circular orbits for a generic stationary axisymmetric metric \cite{2016EL....11630006B} is given
by
\begin{equation}
    \Omega=\frac{d\phi}{dt}=\frac{-\partial_r g_{t\phi}\pm\sqrt{(\partial_r g_{t\phi} )^2-(\partial_r g_{tt})(\partial_r g_{\phi\phi})}}{\partial_r g_{\phi\phi}}
\end{equation}
and for the Hartle-Thorne spacetime it acquires  the  form
\begin{equation}
 \label{eq:velocity}
\Omega=\Omega_0\left[1\mp jW_1(r)+j^2 W_2(r)+qW_3(r)\right],
\end{equation}
where $j=J/M^2$, $q=Q/M^3$,
$\Omega_{0}$ is  the angular velocity  for the Schwarzschild metric, and $W_{1;2;3}$ are reported below 
\begin{eqnarray}
   \Omega_0(r)&=&\frac{M^{1/2}}{r^{3/2}}, \label{eq:Omega0}\\
    W_1(r)&=&\frac{M^{3/2}}{r^{3/2}}, \label{eq:W1} \\
    W_2(r)&=&\left[16M^2r^4(r-2M)\right]^{-1}\Big(48M^7-80M^6r \nonumber\\
          &+&4M^5r^2-18M^4r^3+40M^3r^4+10M^2r^5\nonumber \\
           &+&15Mr^6-15r^7\Big)+W(r) \label{eq:W2} \\
     W_3(r)&=&5\left[16M^2r(r-2M)\right]^{-1}\Big(6M^4-8M^3r\nonumber \\
         &-&2M^2r^2-3Mr^3+3r^4\Big)-W(r),\label{eq:W}\\
     W(r)&=&\frac{15(r^3-2M^3)}{32M^3}\ln\left(\frac{r}{r-2M}\right).
\end{eqnarray}
 
Given the generic orbital angular momentum, 
\begin{equation}
    L=\frac{g_{t\phi}+\Omega g_{\phi\phi}}{\sqrt{-g_{tt}-2\Omega g_{t\phi}-\Omega^2g_{\phi\phi}}},
\end{equation}
we compute it for the Hartle-Thorne metric,  having 
\begin{equation}\label{eq:momentum}
   L= L_0\left[1\mp jH_1(r)+j^2 H_2(r)+q H_3(r)\right],   
\end{equation}
where, analogously to the above case, $L_{0}$, say the  angular momentum  for the Schwarzschild metric and the supporting functions $H_{1;2;3}$ are reported below

\begin{eqnarray}  
 L_{0}(r)&=&r\sqrt{\frac{M}{r-3M}}, \label{eq:L0}\\
      H_{1}(r)&=&\frac{3M^{3/2}(r-2M)}{r^{3/2}(3M-r)},  \label{eq:H1}\\
    H_{2}(r)&=&\left[16M^2r^4(r-3M)^{2}\right]^{-1}\Big[144M^8  \label{eq:H2} \nonumber \\
            &-&144M^7r+20M^6r^2-98M^5r^3 \nonumber \\
            &+&147M^4r^4+205M^3r^5-260M^2r^6\nonumber \\
            &+&105Mr^7-15r^8\Big]+H(r) , \\
    H_{3}(r)&=&5\left[16M^2r(3M-r)\right]^{-1}\Big(6M^4-7M^3r  \label{eq:H3}\nonumber\\
            &-&16M^2r^2+12Mr^3-3r^4\Big)-H(r) ,\\
        H(r)&=&15\left[32M^3(3M-r)\right]^{-1}\Big(6M^4+2M^3r  \label{eq:H}\nonumber\\
            &-&9M^2r^2+5Mr^3-r^4\Big) \ln\left(\frac{r}{r-2M}\right).
\end{eqnarray}

Afterwards, the generic expression for the energy of test particles on circular orbit, 
\begin{equation}
    E=-\frac{g_{tt}+\Omega g_{t\phi}}{\sqrt{-g_{tt}-2\Omega g_{t\phi}-\Omega^2g_{\phi\phi}}},
\end{equation}
leads to 
\begin{equation}
   \label{eq:energy}
E=E_0\left[1\mp jF_{1}(r)+j^2 F_2(r)+q F_3(r)\right],  
\end{equation}
for the Hartle-Thorne spacetime. Here, $E_{0}$ is  the energy for the Schwarzschild space-time and the supporting functions, $F_{1;2;3}$, are
\begin{equation}
 \label{eq:E0}
E_0(r)=\frac{r-2M}{\sqrt{r(r-3M)}}, 
\end{equation}
\begin{equation}
 \label{eq:F1}
F_{1}(r)=\frac{M^{5/2}r^{-1/2}}{(r-2M)(r-3M)}, 
\end{equation}

\begin{eqnarray}
F_2(r)&=&\left[16Mr^4(2M-r)(r-3M)^{2}\right]^{-1} \label{eq:F2}\nonumber\nonumber\\ 
&\times&\Big(-144M^8+144M^7r+28M^6r^2\nonumber\\ 
&+&58M^5r^3+176M^4r^4-685M^3r^5\nonumber\\ 
&+&610M^2r^6-225Mr^7+30r^8\Big)-F(r),   \\
F_{3}(r)&=&-5\left[16Mr(r-2M)(r-3M)\right]^{-1} \label{eq:F3} \nonumber\\
&\times&\Big(6M^4+14M^3r-41M^2r^2\nonumber\\
&+&27Mr^3-6r^4\Big)+F(r),\\
F(r)&=&\frac{15r(8M^2-7Mr+2r^2)}{32M^2(3M-r)} \ln\left(\frac{r}{r-2M}\right) . \quad  \label{eq:F}
\end{eqnarray}

As mentioned in the previous section, these quantities reduce to the corresponding Schwarzschild values for  $q\to 0$ and $j\to 0$. 

\begin{figure*}[ht]
\begin{minipage}{0.49\linewidth}
\center{\includegraphics[width=0.97\linewidth]{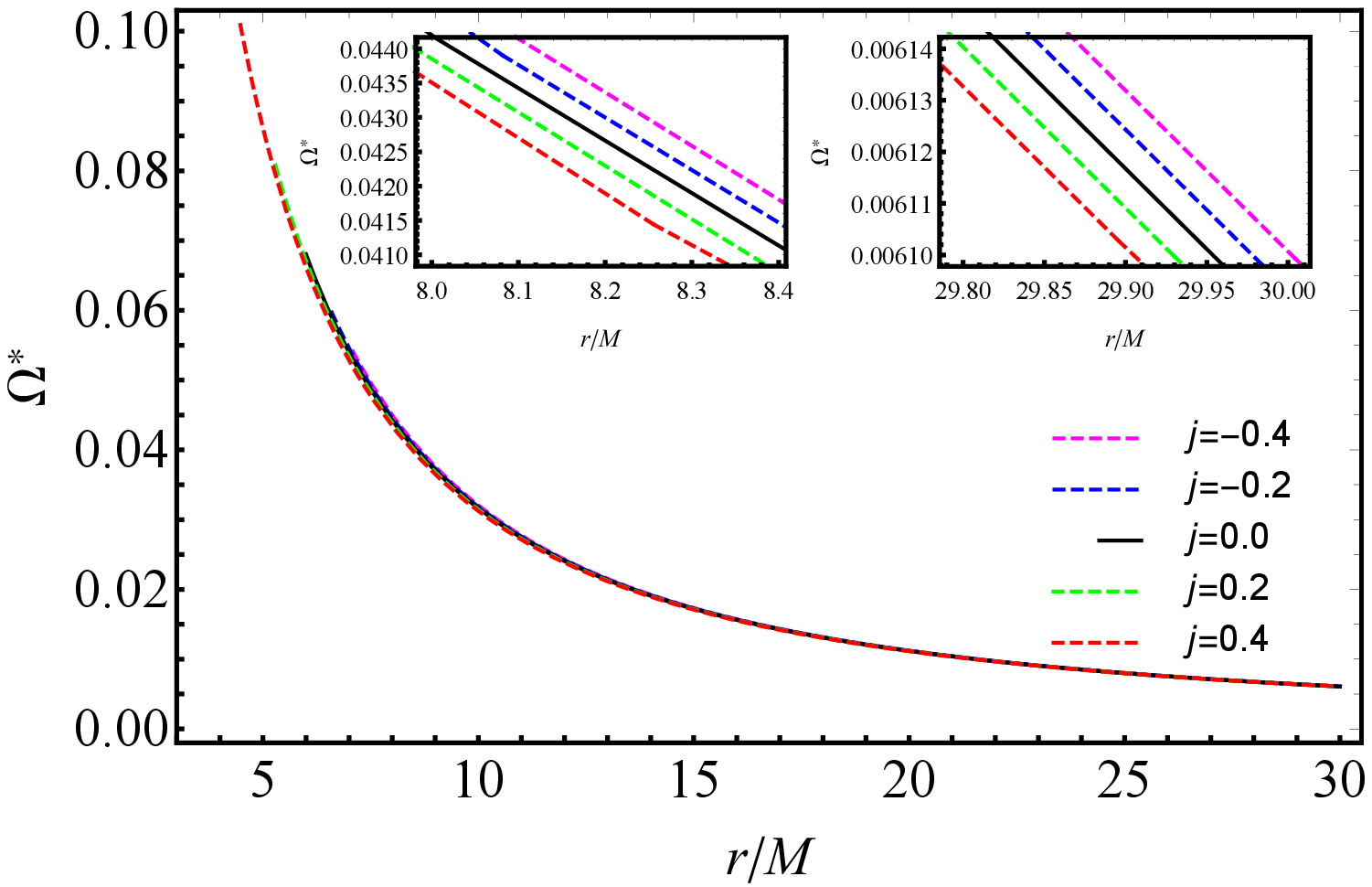}\\ }
\end{minipage}
\hfill 
\begin{minipage}{0.50\linewidth}
\center{\includegraphics[width=0.97\linewidth]{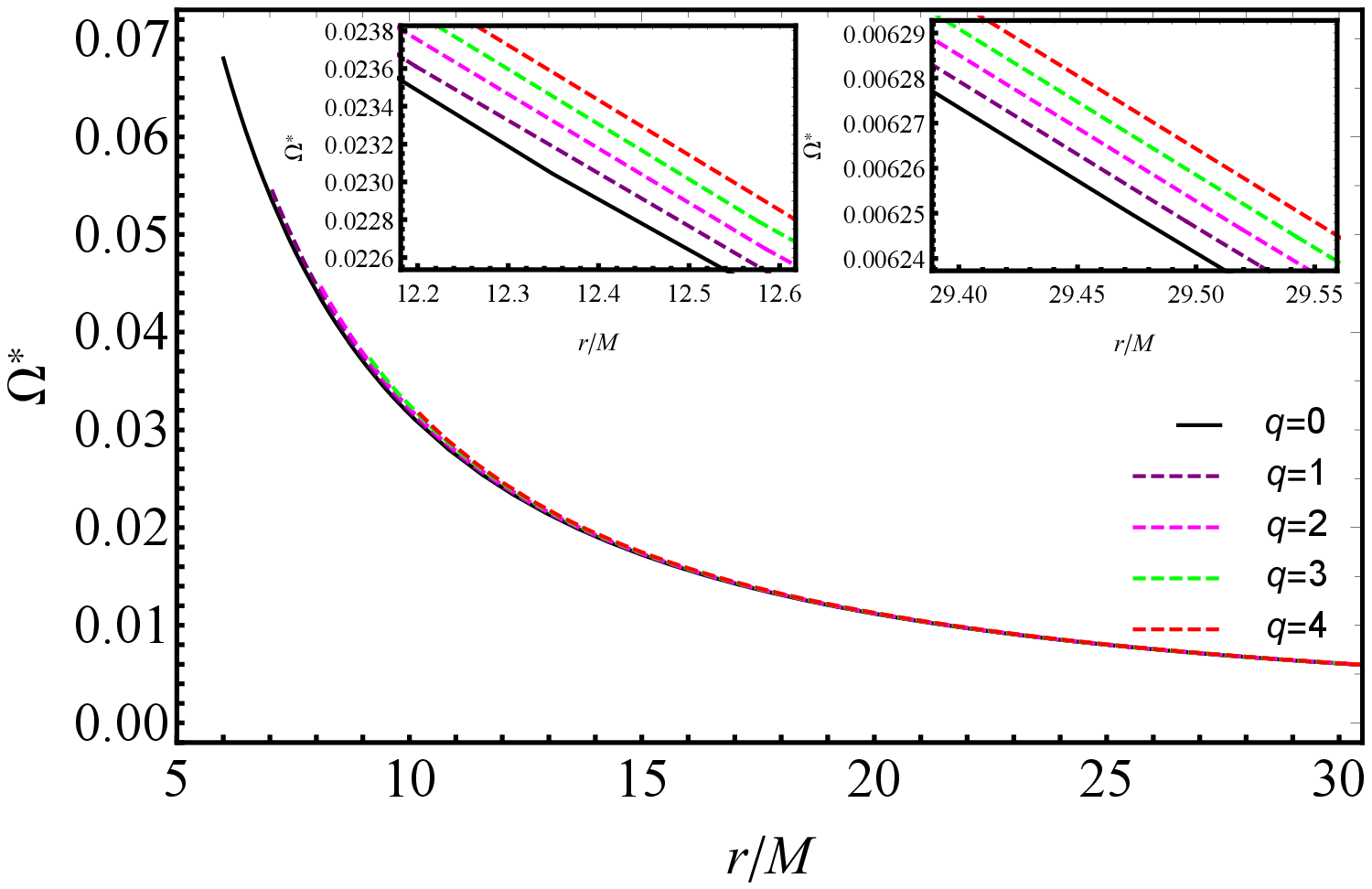}\\ }
\end{minipage}
\caption{Dimensionless angular velocity $\Omega^*=M \Omega$ of test particles versus radial distance $r$ normalized by the total mass $M$ in the Hartle-Thorne spacetime. All curves start from $R_{ISCO}$. Left panel: the quadrupole parameter is set as $q=0$ in all curves. Right panel: the spin parameter is set as $j=0$ in all curves.}
\label{fig:omega}
\end{figure*}
\begin{figure*}[ht]
\begin{minipage}{0.49\linewidth}
\center{\includegraphics[width=0.97\linewidth]{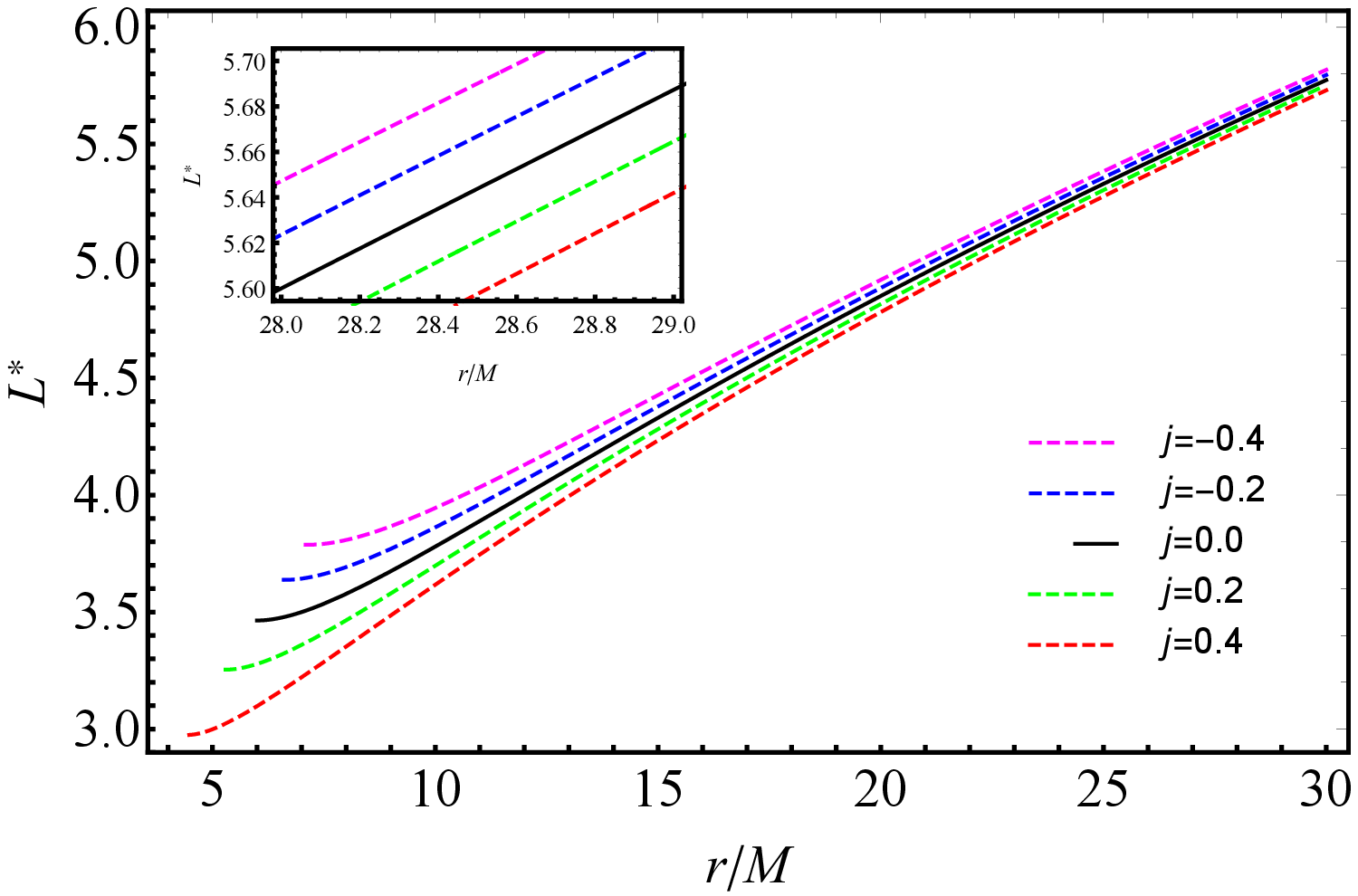}\\ } 
\end{minipage}
\hfill 
\begin{minipage}{0.50\linewidth}
\center{\includegraphics[width=0.97\linewidth]{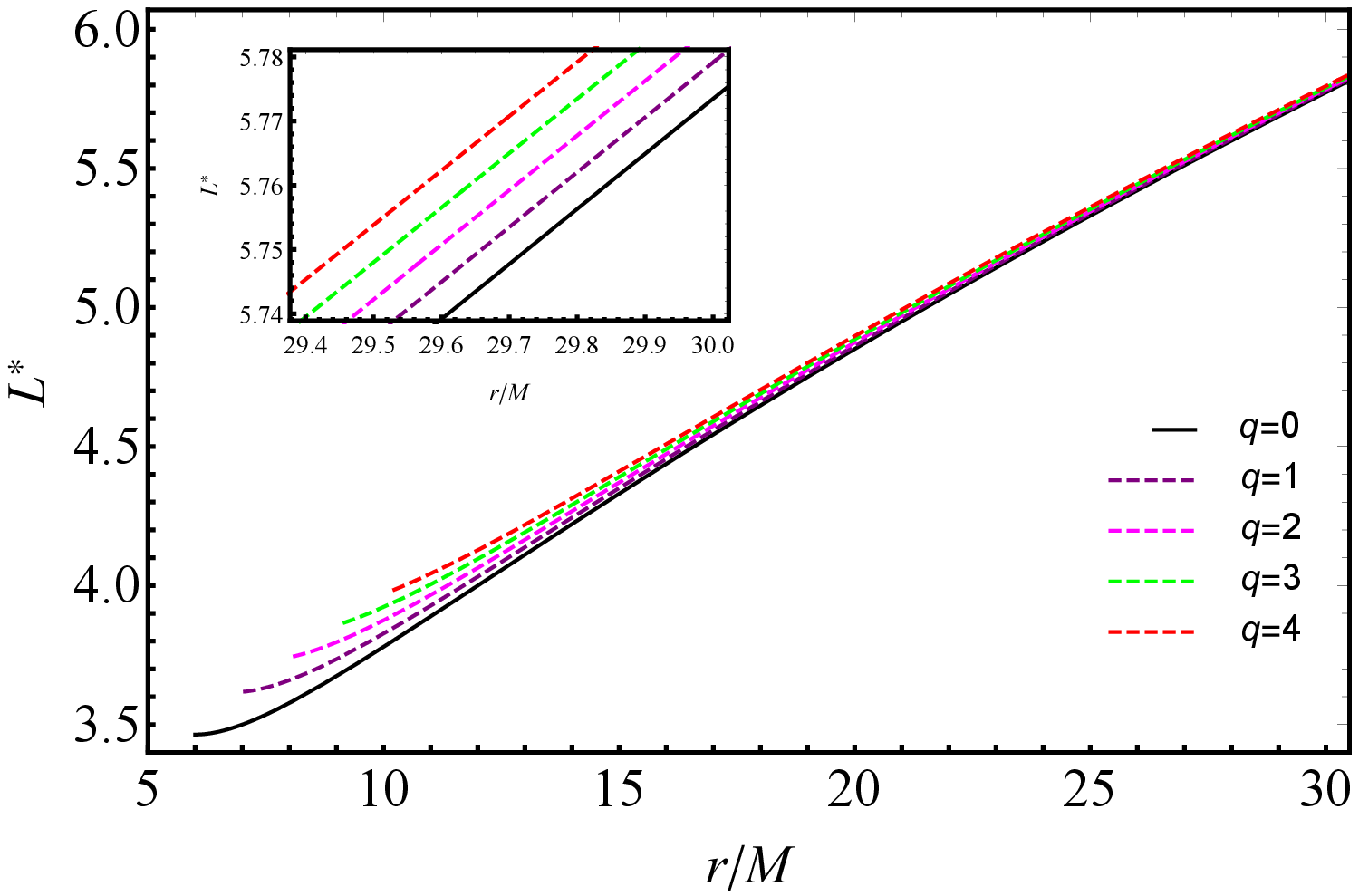}\\ }
\end{minipage}
\caption{Angular momentum $L^*=L/M$ of test particles versus radial distance $r$ normalized by the total mass $M$ in the Hartle-Thorne metric. All curves start from $R_{ISCO}$. Left panel: $L^*$ for fixed $q=0$ and arbitrary $j$. Right panel: for fixed $j=0$ and arbitrary $q$.}
\label{fig:angmom}
\end{figure*}
\begin{figure*}[ht]
\begin{minipage}{0.49\linewidth}
\center{\includegraphics[width=0.97\linewidth]{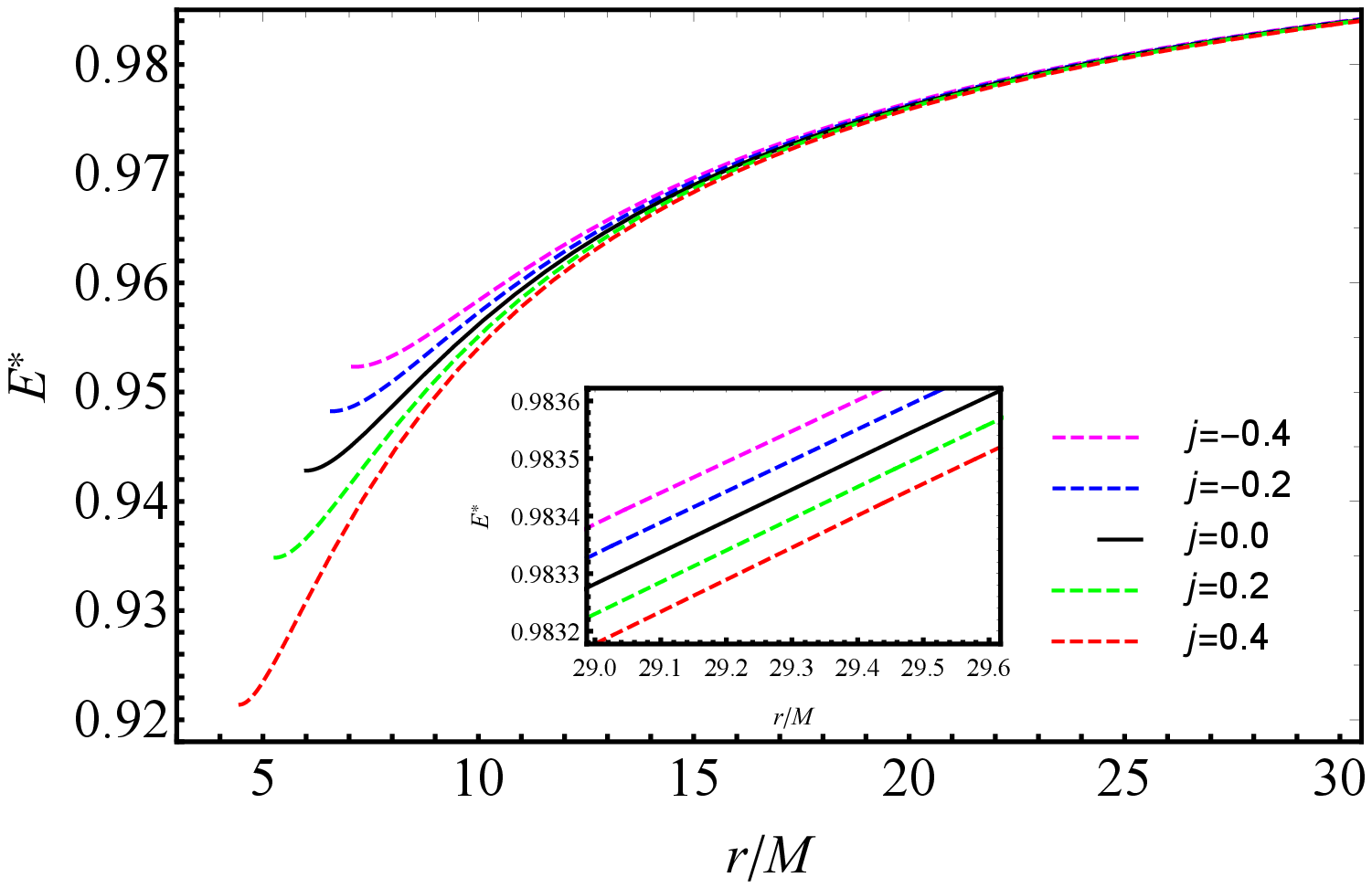}\\ } 
\end{minipage}
\hfill 
\begin{minipage}{0.50\linewidth}
\center{\includegraphics[width=0.97\linewidth]{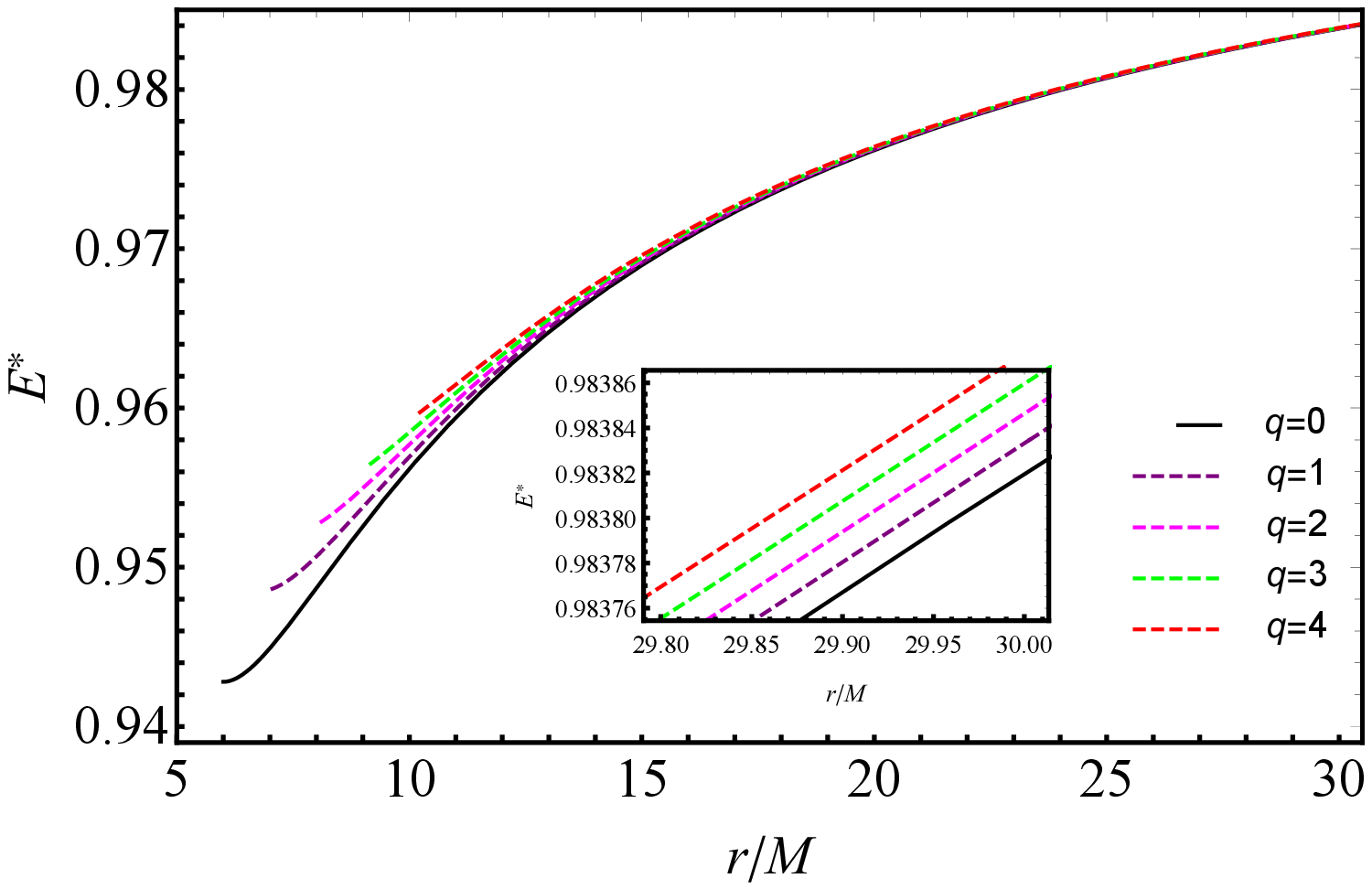}\\ }
\end{minipage}
\caption{Energy $E^*=E$ of test particles versus radial distance $r$ normalized by total mass $M$ in the Hartle-Thorne spacetime. All curves start from $R_{ISCO}$. Left panel: $E^*$ for fixed $q=0$ and arbitrary $j$. Right panel: $E^*$ for fixed $j=0$ and arbitrary $q$.}
\label{fig:energy}
\end{figure*}
%

\subsection{The innermost stable circular orbits}
As we claimed in the introduction, ISCO is the smallest circular orbit around a massive object that can be obtained within the contexts of metric theories. Computing this quantity is manifestly relevant since, there,  a test particle can maintain a stable orbit, and so the ISCO radius is function of the mass and angular momentum of the central object determining the accretor. Consequently, in the context of black hole accretion disks, the ISCO is of utmost importance as it represents the inner boundary of the disk itself. Thus, the ISCO radius, dubbed $R_{ISCO}$, is defined via the condition $dL/dr=0$ or $dE/dr=0$ \cite{2016PhRvD..93b4024B} and is given by
\begin{align}
 \label{eq:risco}
 \nonumber
R_{ISCO}&=6M\left[1\mp\frac{2}{3}\sqrt{\frac{2}{3}}j+\left(\frac{251647}{2592}-240\ln\frac{3}{2}\right)j^2\right. \\
&\left.+\left(-\frac{9325}{96}+240\ln\frac{3}{2}\right)q\right] \nonumber \\
&\approx 6M[1\mp0.5443j-0.2256j^2+0.1762q]\,.
\end{align}
Here, the sign in front of $j$ in the expression for the ISCO radius determines the rotation direction of the central object. For co-rotating or prograde orbits, a negative sign is used, while for counter-rotating or retrograde orbits, a positive sign is used. It is worth noting that in the limit of zero angular momentum and quadrupole moment, the ISCO radius reduces to $R_{\text{ISCO}}^0=6M$ for the Schwarzschild spacetime.

In addition, one of the key quantities which is of great interest is the \emph{efficiency of converting matter into radiation} (see for details page 662 of Ref.~\cite{1973grav.book.....M})
\begin{equation}\label{eq:efficiency}
    \eta=[1-E(R_{ISCO})]\times100\% ,
\end{equation}
where the energy of test particles $E$ is calculated at $R_{ISCO}$.

\subsection{Numerical analysis of angular velocity, angular momentum and energy of test particles}
Here, we report an overall analysis of our numerical findings concerning the angular velocity, angular momentum, the energy of test particles in circular orbits, and all the kinematic quantities of our metric, which are useful for the study of the accretion disk.

Specifically, in Fig.~\ref{fig:omega}, we present the orbital angular velocity $\Omega^*(r)=M \Omega$ of test particles as a function of the normalized radial coordinate $r/M$ in the Hartle-Thorne metric. In the left panel, we consider a fixed $q=0$ and arbitrary $j$, while in the right panel, we fix $j=0$ and vary $q$. In the left panel, the curves with $j>0$ are similar to the co-rotating orbits in the Kerr metric or the $q$-metric with $q>0$, we observe that the angular velocity curves lie below the curve corresponding to the Schwarzschild metric. Instead the curves with $j<0$ correspond to the counter-rotating orbits in the Kerr spacetime and lie above the curve corresponding to the Schwarzschild metric. In the right panel, which resembles the counter-rotating orbits in the Kerr metric, we see that the angular velocity curves are above the curve obtained in the Schwarzschild solution. It is important to note that in the Hartle-Thorne spacetime, $q>0$ corresponds to oblate astrophysical objects, while $q<0$ corresponds to prolate objects. Since most rotating objects are oblate, we focus on cases where $q>0$ throughout this paper. Furthermore, the limit $q=0$ describes rotating objects without deformation, distinct from the Kerr spacetime, although many features are similar to the Kerr metric. On the other hand, the limit $j=0$ represents static deformed objects.

In Fig.~\ref{fig:angmom}, we present the dimensionless orbital angular momentum $L^*(r)=L/M$ of test particles as a function of the normalized radial coordinate $r/M$ in the Hartle-Thorne metric. The left panel corresponds to fixed $q=0$ and arbitrary $j$, while the right panel represents fixed $j=0$ and arbitrary $q$. In the left panel, it can be observed that the curves of $L^{*}$ for $j>0$ ($j<0$) lie below (above) with respect to the Schwarzschild case. Moreover,  the overall behavior of $L^{*}$ resembles that of co-rotating (counter-rotating) test particles' orbits in the Kerr spacetime. On the right panel, the Schwarzschild case exhibits smaller values of $L^{*}$, and the general trend of the curves is similar to the ones observed in the Kerr metric for counter-rotating orbits or the $q$-metric with $q>0$ cases (see \cite{2021PhRvD.104h4009B} for more detailed information). 

In Fig.~\ref{fig:energy}, we depict the energy per unit mass $E^*=E$ of test particles as a function of the normalized radial coordinate $r/M$ in the Hartle-Thorne metric. The left panel corresponds to fixed $q=0$ and arbitrary $j$, while the right panel represents fixed $j=0$ and arbitrary $q$. In the left panel, it is evident that the energy of particles for $j>0$ ($j<0$) cases is relatively lower (higher) than in the Schwarzschild case. We observe a similarity between the behavior of $E^*$ for co-rotating and counter-rotating orbits in the Kerr spacetime \cite{2021PhRvD.104h4009B}. On the right panel, the curve of $E^*$ for the Schwarzschild case is below the curves of $q>0$ cases, resembling the behavior observed in the $q$-metric with $q>0$ case or the Kerr metric with counter-rotating orbits \cite{2021PhRvD.104h4009B}.
\begin{figure*}[ht]
\begin{minipage}{0.48\linewidth}
\center{\includegraphics[width=1.02\linewidth]{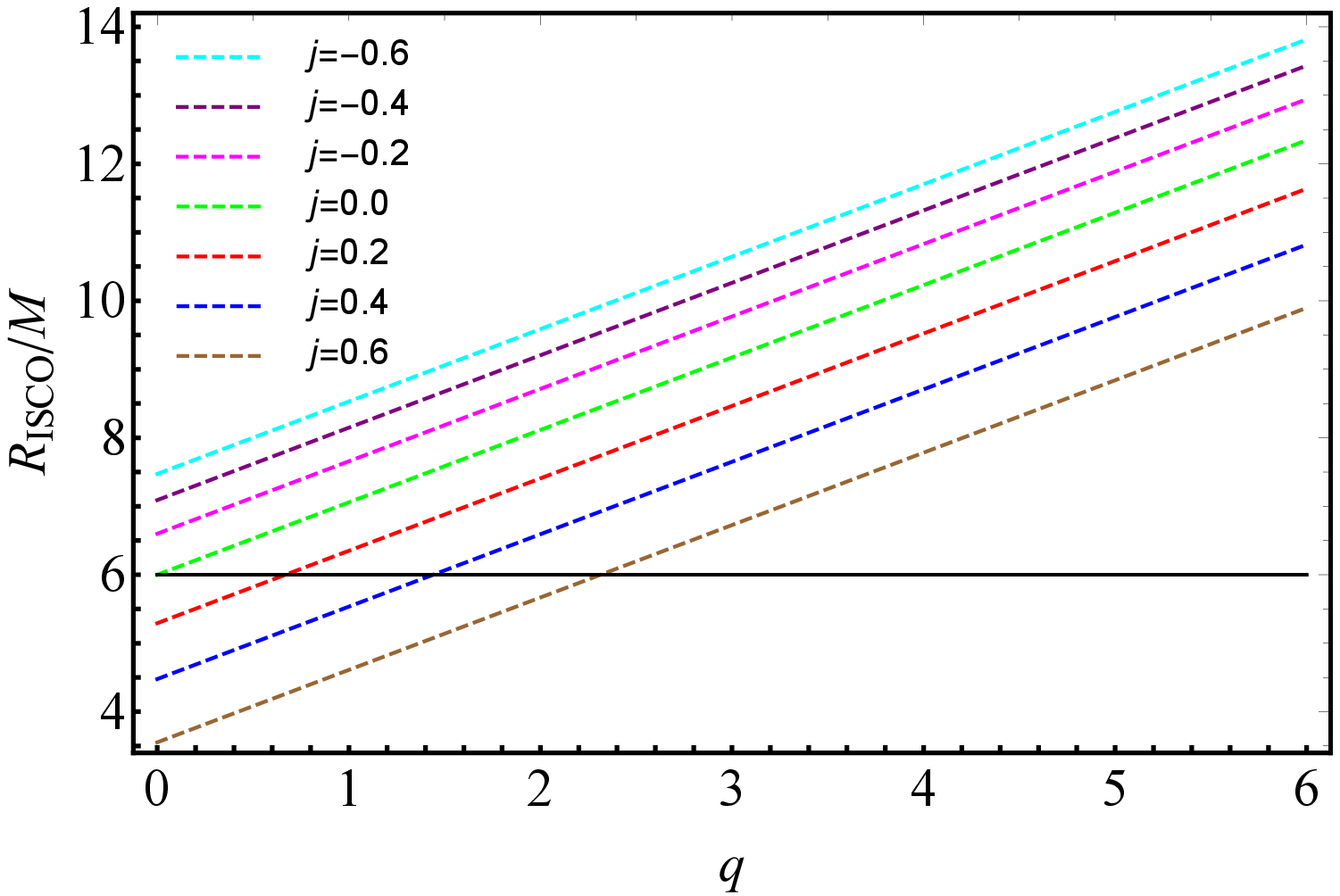}\\ }
\end{minipage}
\hfill 
\begin{minipage}{0.50\linewidth}
\center{\includegraphics[width=0.97\linewidth]{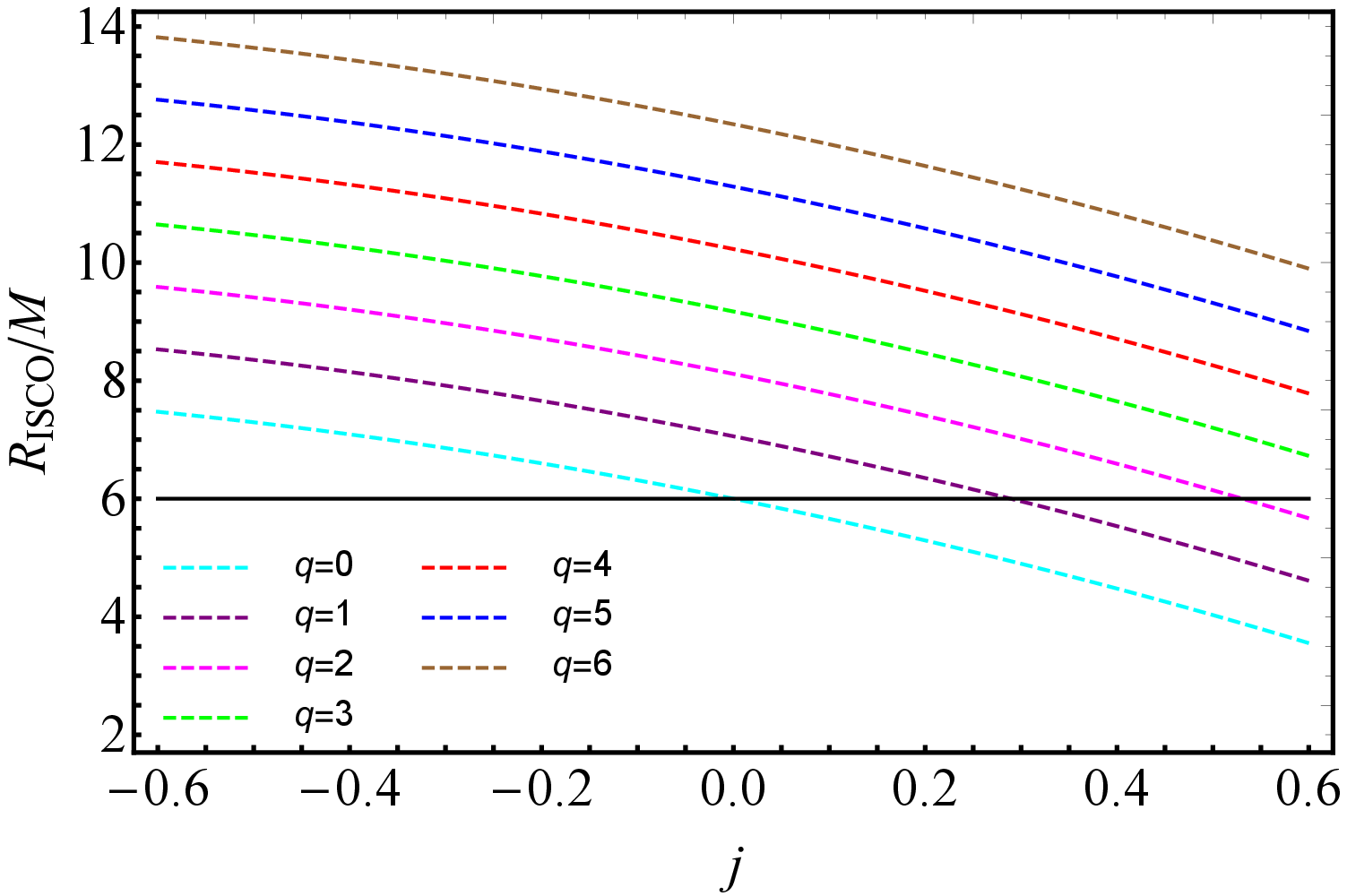}\\ }
\end{minipage}
\caption{$R_{ISCO}$ in the Hartle-Thorne spacetime. Left panel:  $R_{ISCO}$ versus $q$ for different values of $j$. Right panel: $R_{ISCO}$ versus $j$ for different values of $q$.}
\label{fig:Riscoqj}
\end{figure*}
In Fig.~\ref{fig:Riscoqj}, we illustrate the dependence of the normalized ISCO radius $R_{ISCO}/M$ in terms of $j$ and $q$. The left panel shows $R_{ISCO}/M$ versus $q$ for fixed values of $j$. Here, counter-rotating orbits will have larger $R_{ISCO}/M$ with respect to the Schwarzschild (black solid straight line) for increasing $q$. However, co-rotating orbits will have smaller $R_{ISCO}/M$ with respect to the Schwarzschild case (black solid straight line) for decreasing $q$. The right panel presents $R_{ISCO}/M$ versus $j$ for fixed values of $q$. Counter-rotating orbits will have larger $R_{ISCO}/M$ with respect to the Schwarzschild (solid straight line) for any value of $q$, but for co-rotating orbits $R_{ISCO}/M$ will be smaller with respect to the Schwarzschild case for small values of $q$. The two panels complement each other. One can see that $R_{ISCO}/M$ in the field of rotating neutron stars with large $j$ can be smaller than in the field of a static neutron star or a Schwarzschild black hole with equal masses. This tendency can be explained by the fact that for small values of $q$ the effects of $j$ start prevailing and the Hartle-Thorne metric behaves like the Kerr metric. Correspondingly, one can observe a similar effect for both co-rotating and counter-rotating orbits in the Kerr spacetime.
\begin{figure*}[ht]
\begin{minipage}{0.48\linewidth}
\center{\includegraphics[width=1.02\linewidth]{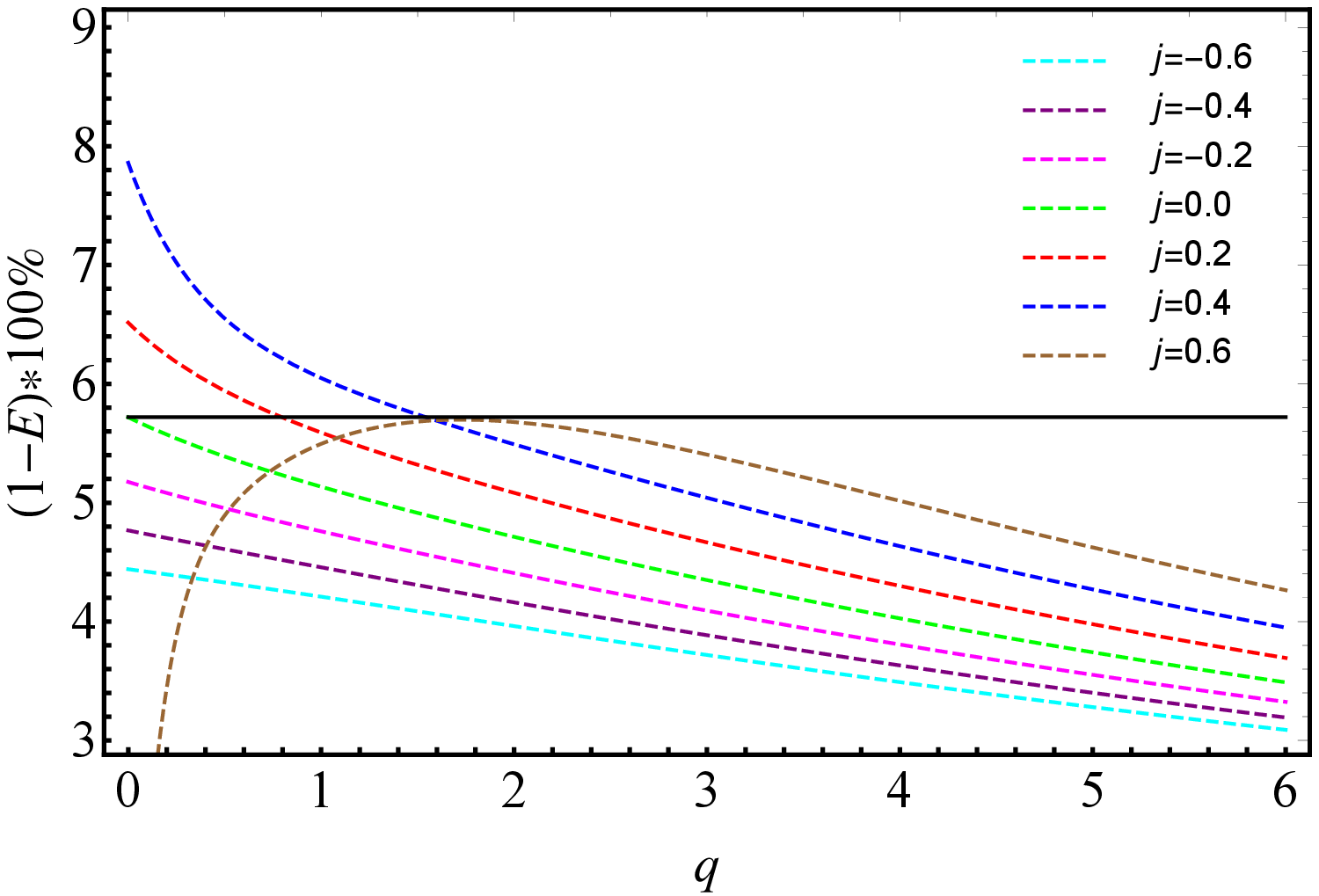}\\ }
\end{minipage}
\hfill 
\begin{minipage}{0.50\linewidth}
\center{\includegraphics[width=0.97\linewidth]{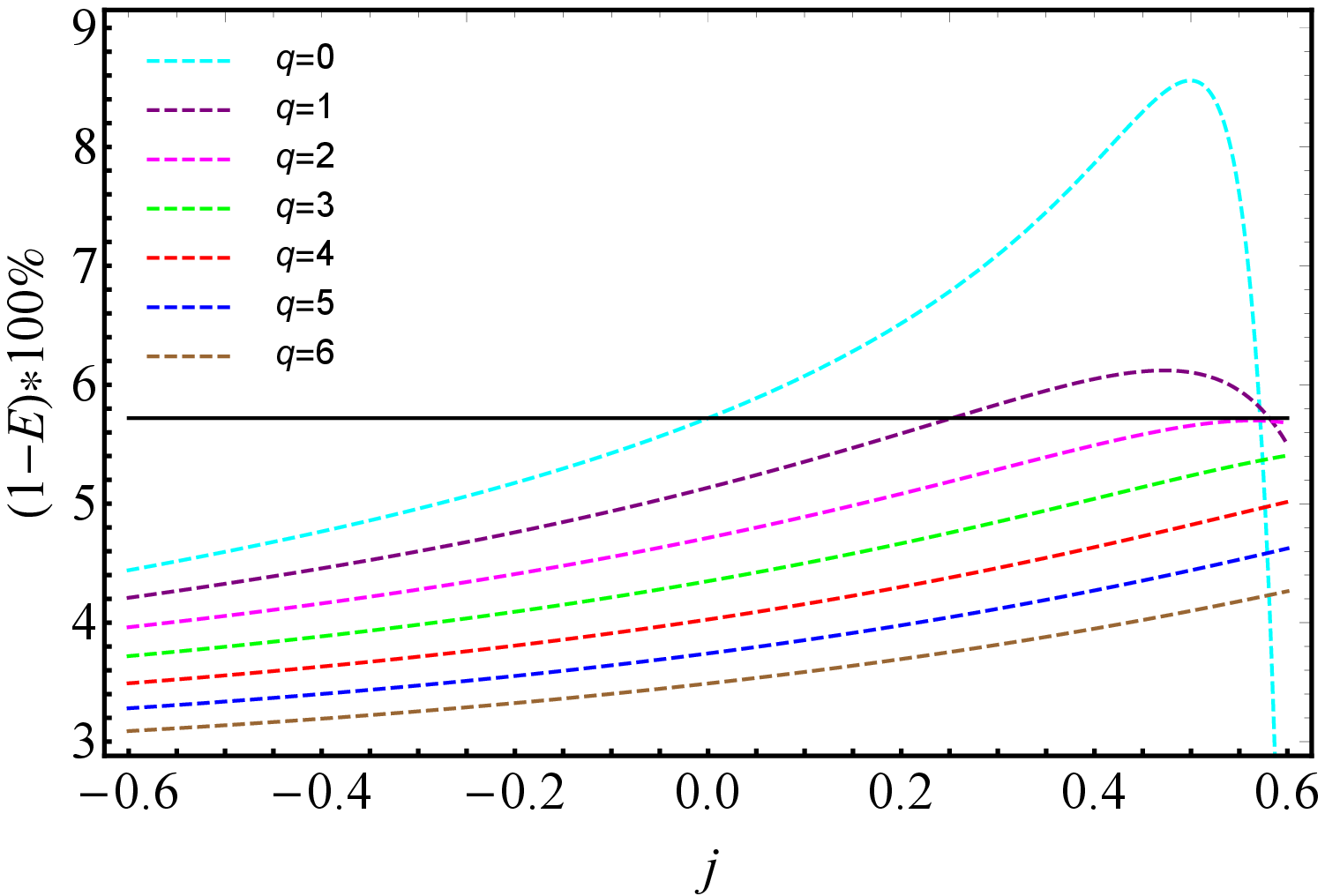}\\ }
\end{minipage}
\caption{Efficiency $\eta$ of converting matter into radiation in the Hartle-Thorne spacetime. Left panel: $\eta$ versus $q$ for different values of $j$. Right panel:  $\eta$ versus $j$ for different values of $q$.}
\label{fig:Efficiencyqj}
\end{figure*}

In Fig.~\ref{fig:Efficiencyqj}, the efficiency of the compact object is illustrated in the Hartle-Thorne spacetime. Left panel shows that for increasing $q$ the efficiency decreases  for different values of $j$ and starting from certain values of $q$ the efficiency becomes smaller than in the Schwarzschild case (black solid straight line). Instead, for smaller values of $q$ the efficiency of neutron stars for co-rotating orbits can be larger than the one of a static neutral black hole, possessing the same mass. Here again as in a previous case for small $q$ the effects of $j$ are more dominant. Therefore the trend of the curves are similar to the ones of the Kerr metric for co-rotating orbits. Right panel shows efficiency versus $j$ for $q\geq0$. As expected for small $q$ and large $j$ the efficiency will be larger than for the Schwarzschild black hole (black solid straight line).

\section{Spectra of thin accretion disks}\label{sez4}
To investigate the luminosity and spectral characteristics of the accretion disk in the Hartle-Thorne spacetime, we adopt the simplest model for accretion disk, developed by Novikov-Thorne and Page-Thorne as described in Refs. \cite{novikov1973, page1974}. The underlying relativistic models for accretion disks around black holes are limited in their validity and physical realism at the inner edge due to the boundary condition that requires a sudden cessation of viscous stresses at the radius separating the region of circular orbits from the region of spiral orbits. 

However, in the case of accreting black holes, emission from the inner disk provides insights into the corresponding black hole spin. According to the model, for a thin accretion disk around a black hole, the accreting matter gradually moves inward along nearly Keplerian orbits due to viscous evolution until it reaches the radius of the innermost stable circular orbit (ISCO), beyond which the gas rapidly falls into the black hole. 

Consequently, the inner edge of the viscous accretion disk is predicted to be very close to the ISCO. In this respect, most analytical models of accretion disks assume a stationary and axially symmetric state of the matter being accreted into the black hole. In these scenarios, all physical quantities depend only on two spatial coordinates: the radial distance from the center  and the vertical distance from the equatorial symmetry plane. 

Even though commonly studied models assume that the disk is not significantly vertically extended, the Novikov-Thorne solution represents local solutions and introduces an assumption that the viscous torque vanishes at the ISCO, leading to a singularity in the model at that point. 

For very low accretion rates, this singularity does not significantly affect the electromagnetic spectrum or several other important astrophysical predictions of the model\footnote{However, in certain astrophysical scenarios where the inner boundary condition plays a crucial role, such as global modes of disk oscillations, the Novikov-Thorne model is inadequate.}. 

\begin{figure*}[ht]
\begin{minipage}{0.49\linewidth}
\center{\includegraphics[width=0.97\linewidth]{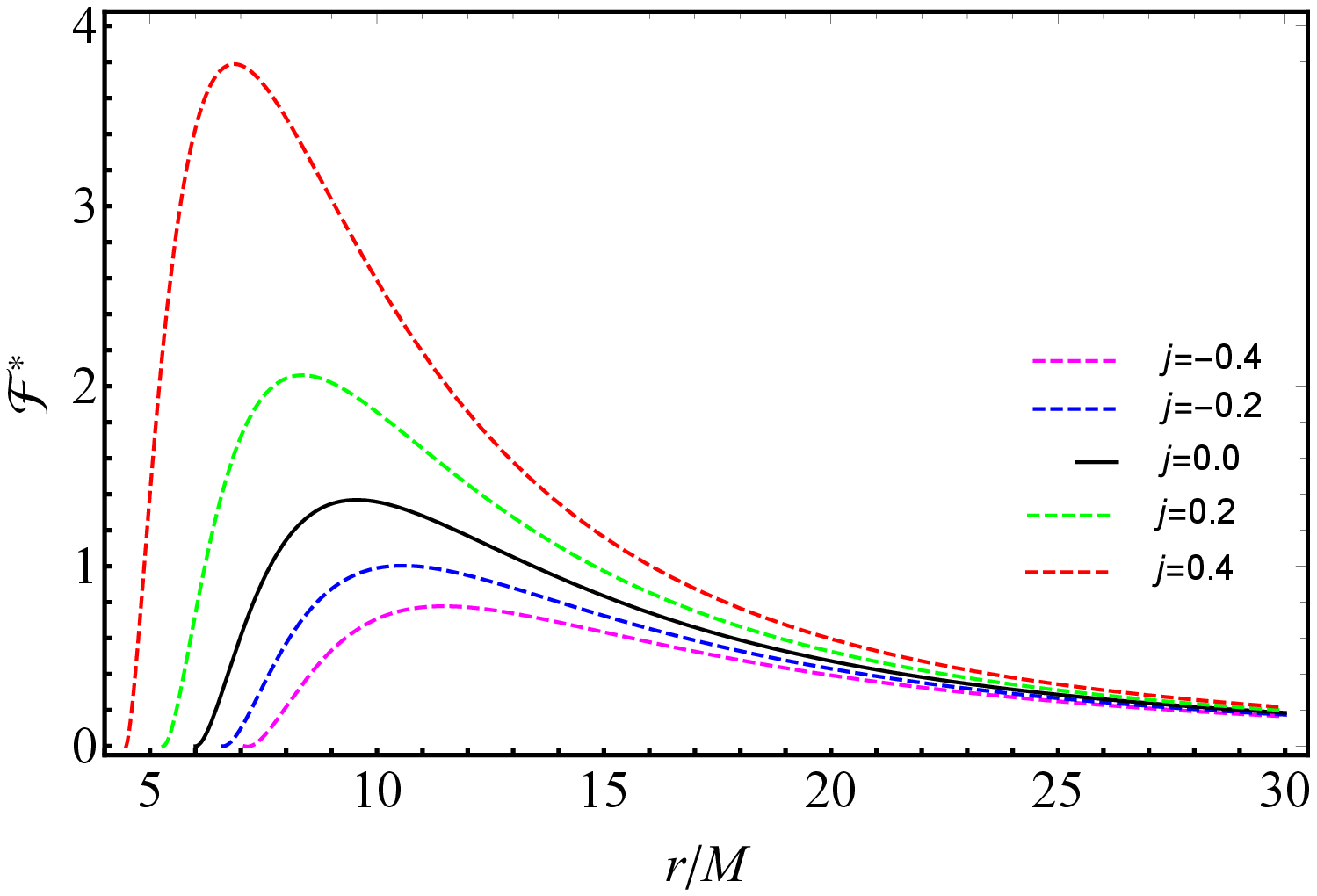}\\ } 
\end{minipage}
\hfill 
\begin{minipage}{0.50\linewidth}
\center{\includegraphics[width=0.97\linewidth]{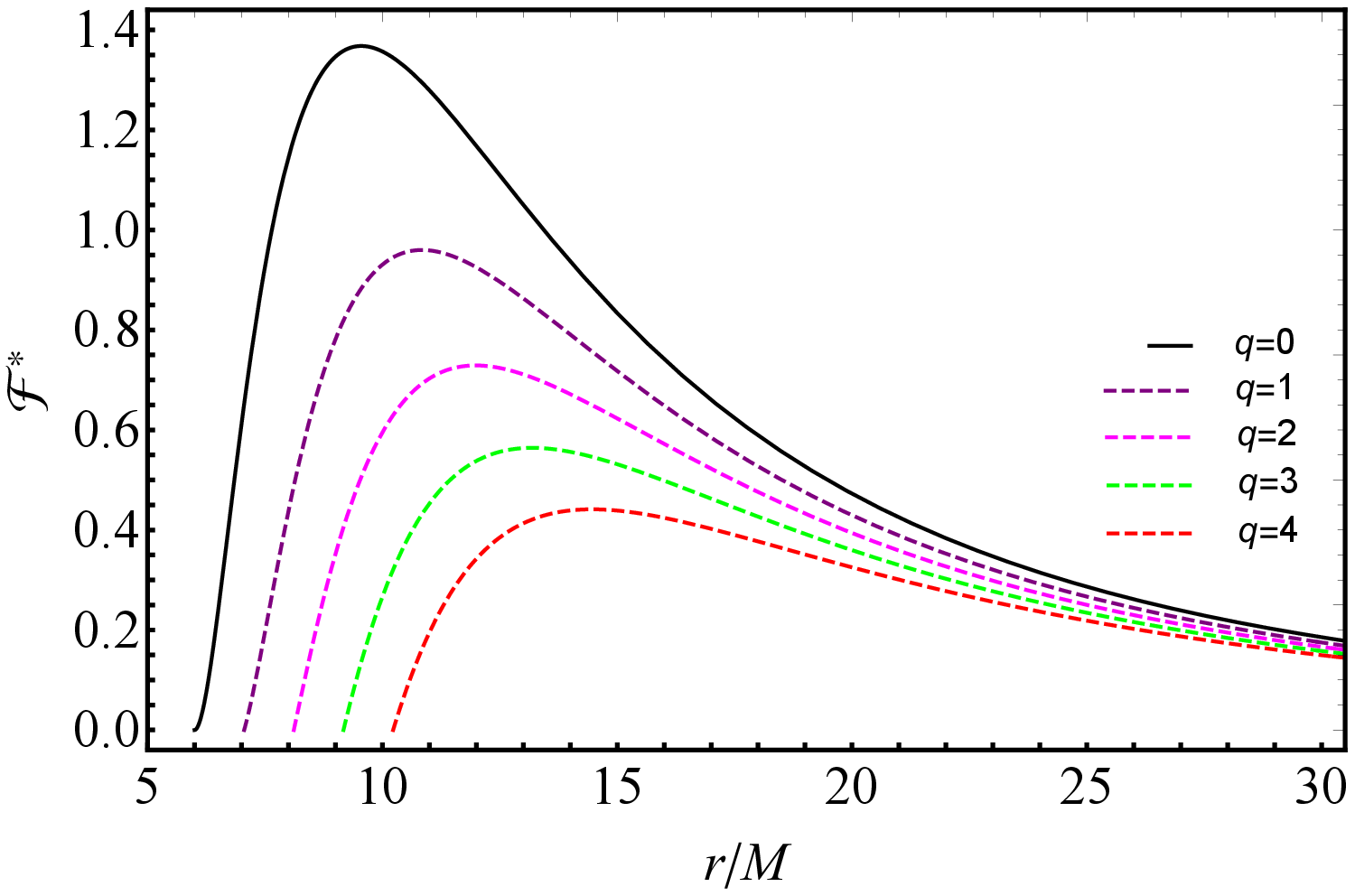}\\ }
\end{minipage}
\caption{Radiative flux $\mathcal{F}^*$ multiplied by $10^5$ of the accretion disk in the Hartle-Thorne metric versus radial distance $r$ normalized by the total mass $M$. Left panel: $\mathcal{F}^*$ versus $r/M$ for fixed $q=0$ and arbitrary $j$. Right panel: $\mathcal{F}^*$ versus $r/M$ for fixed $j=0$ and arbitrary $q\geq0$. } 
\label{fig:flux}
\end{figure*}

In the subsequent analysis, we make the assumption of a constant mass accretion rate of the disk to meet the aforementioned criteria. Therefore, the primary quantities that describe the model locally are the radiative flux $\mathcal{F}$ and the differential luminosity, which represents the energy per unit time reaching an observer located at infinity, denoted as $\mathcal{L}_{\infty}$. The differential luminosity $\mathcal{L}_{\infty}$ is estimated based on the flux $\mathcal{F}$, which represents the energy radiated per unit area per unit time by the accretion disk. We thus have\footnote{The flux definition can be modified conveniently. Later, we take its normalized version with respect to total mass $M$, say $\mathcal{F}^*(r)=M^2\mathcal{F}(r)$. Suitably this will help in formulating spectral properties of the disk.} 
\cite{novikov1973, page1974}

\begin{equation}\label{eq:flux}
\mathcal{F}(r)=-\frac{\dot{{\rm m}}}{4\pi \sqrt{-g}} \frac{\Omega_{,r}}{\left(E-\Omega L\right)^2 }\int^r_{r_{i}} \left(E-\Omega L\right) L_{,\tilde{r}}d\tilde{r},
\end{equation}
and 
\begin{equation}
 \label{eq:difflum}
\frac{d\mathcal{L}_{\infty}}{d\ln{r}}=4\pi r \sqrt{-g}E \mathcal{F}(r),
\end{equation}
where in the previous discussion,  the term flux refers to the energy radiated per unit area per unit time by the accretion disk.

In the given expression, $r_i=R_{ISCO}$ represents the value of the radial coordinate corresponding to the innermost stable circular orbit (ISCO), while $\dot{{\rm m}}$ denotes the mass accretion rate of the disk, which is assumed to be constant throughout the analysis. Lastly, $g$ represents the determinant of the metric tensor in the three-dimensional subspace spanned by the coordinates $(t, r, \varphi)$. In the above relation, the effect of the geometry enters through $g$, namely the determinant of the metric of the three-dimensional sub-space ($t,r,\varphi$) (i.e. $\sqrt{-g}=\sqrt{-g_{rr}(g_{tt}g_{\varphi\varphi}-g_{t\varphi}^2}$) \cite{2012ApJ...761..174B}.

\subsection{Emitting properties}
The radiative flux and its local definition at infinity represent the radiation emitted by the disk as a function of the radial coordinate. However, they do not consider the measured quantities that involve the spectrum of light and its frequencies.

In practice, we observe the emitted spectrum as a function of frequency. Therefore, it is natural to consider the determination of the spectral luminosity distribution observed at infinity, denoted as $\mathcal{L}_{\nu,\infty}$. Assuming that the overall emission follows a black body radiation pattern and considering $u^t$, the contra-variant time component of the four velocity and $\nu$, the frequency of the emitted radiation, we can handle for the $\mathcal{L}_{\nu,\infty}$ the following formula \cite{2020MNRAS.496.1115B}:
\begin{equation}\label{eq:speclum}
\nu \mathcal{L}_{\nu,\infty}=\frac{60}{\pi^3}\int^{\infty}_{r_{i}}\frac{\sqrt{-g }E}{M^2}\frac{(u^t y)^4}{\exp\left[u^t y/\mathcal{F}^{*1/4}\right]-1}dr,
\end{equation}
with the positions
\begin{subequations}
\begin{align}\label{eq:sample7}
u^t(r)&=\frac{1}{\sqrt{-g_{tt}-2\Omega g_{t\varphi}-\Omega^2 g_{\varphi \varphi}}},\\
y&=h\nu/kT_*\,.
\end{align}
\end{subequations}

Besides the clear definitions of $h$ and $k$ as  Planck's and Boltzmann's constants respectively, the relation \eqref{eq:speclum} is also function of the characteristic temperature, $T_*$. 
This quantity can be obtained adopting the aforementioned approximation of black body. 
Hence, taking into account the above Stefan-Boltzmann law we immediately obtain 
\begin{equation}
    \sigma T_*^4=\frac{1}{4\pi}\frac{\dot{{\rm m}}}{ M^2}\,,
\end{equation}

\begin{figure*}[ht]
\begin{minipage}{0.49\linewidth}
\center{\includegraphics[width=0.97\linewidth]{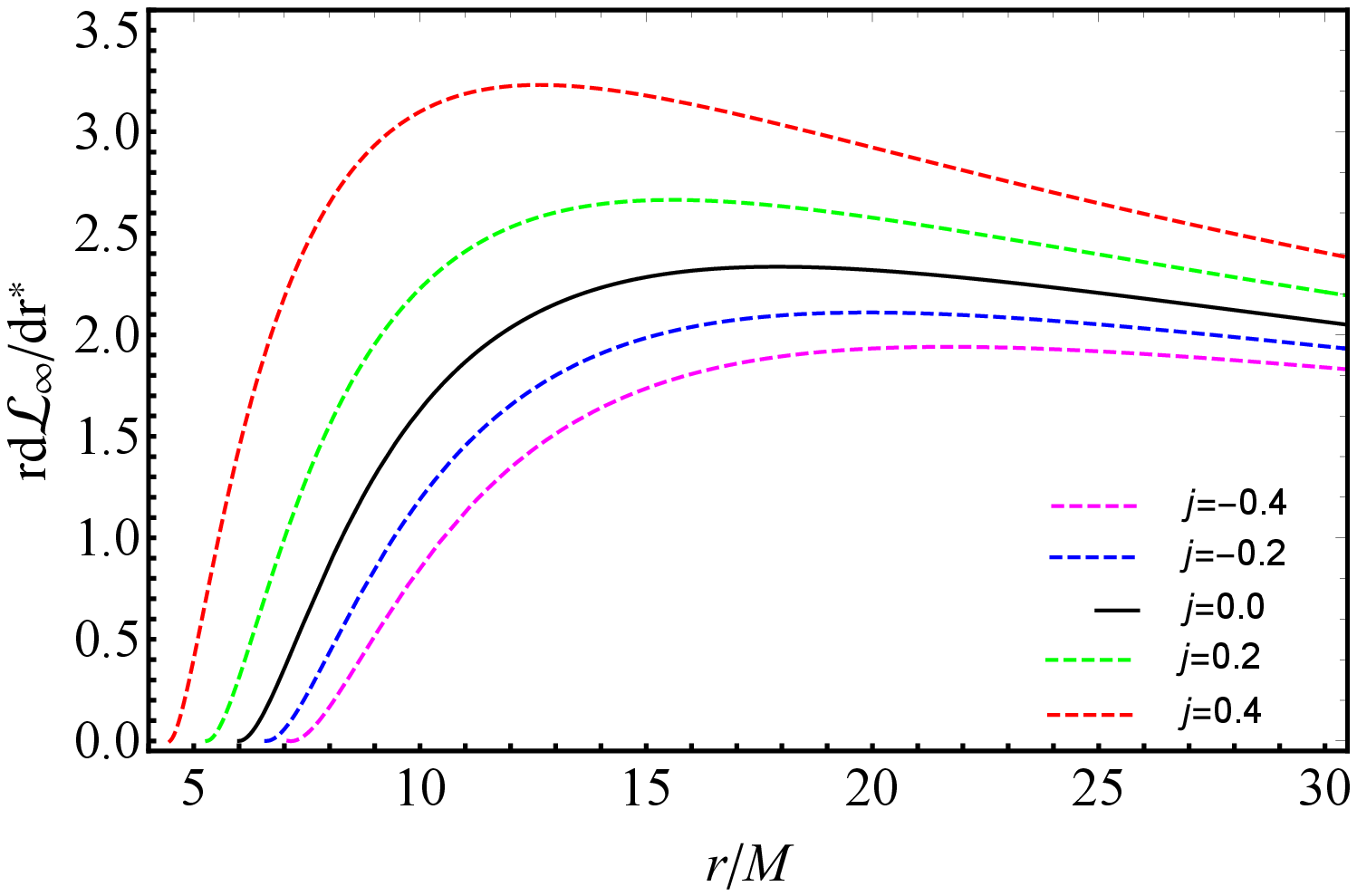}\\ } 
\end{minipage}
\hfill 
\begin{minipage}{0.50\linewidth}
\center{\includegraphics[width=0.97\linewidth]{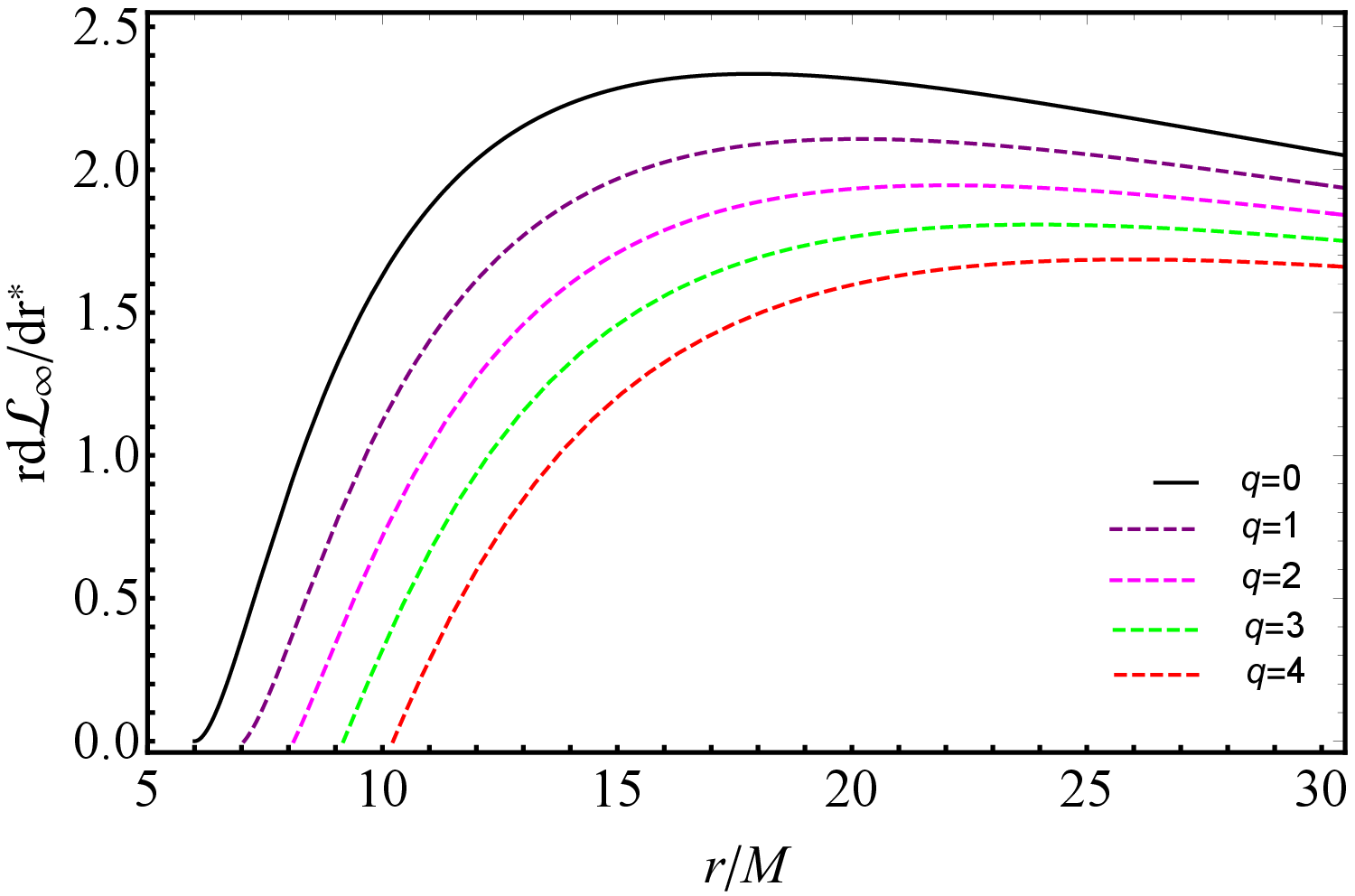}\\ }
\end{minipage}
\caption{Differential luminosity multiplied by $10^2$ of the accretion disk in the Hartle-Thorne metric versus radial coordinate $r$ normalized by total mass $M$. Left panel: Differential luminosity versus $r/M$ for fixed $q=0$ and arbitrary $j$. Right panel: Differential luminosity versus $r/M$ for fixed $j=0$ and arbitrary $q\geq0$.}
\label{fig:difflum}
\end{figure*}
\begin{figure*}[ht]
\begin{minipage}{0.49\linewidth}
\center{\includegraphics[width=0.97\linewidth]{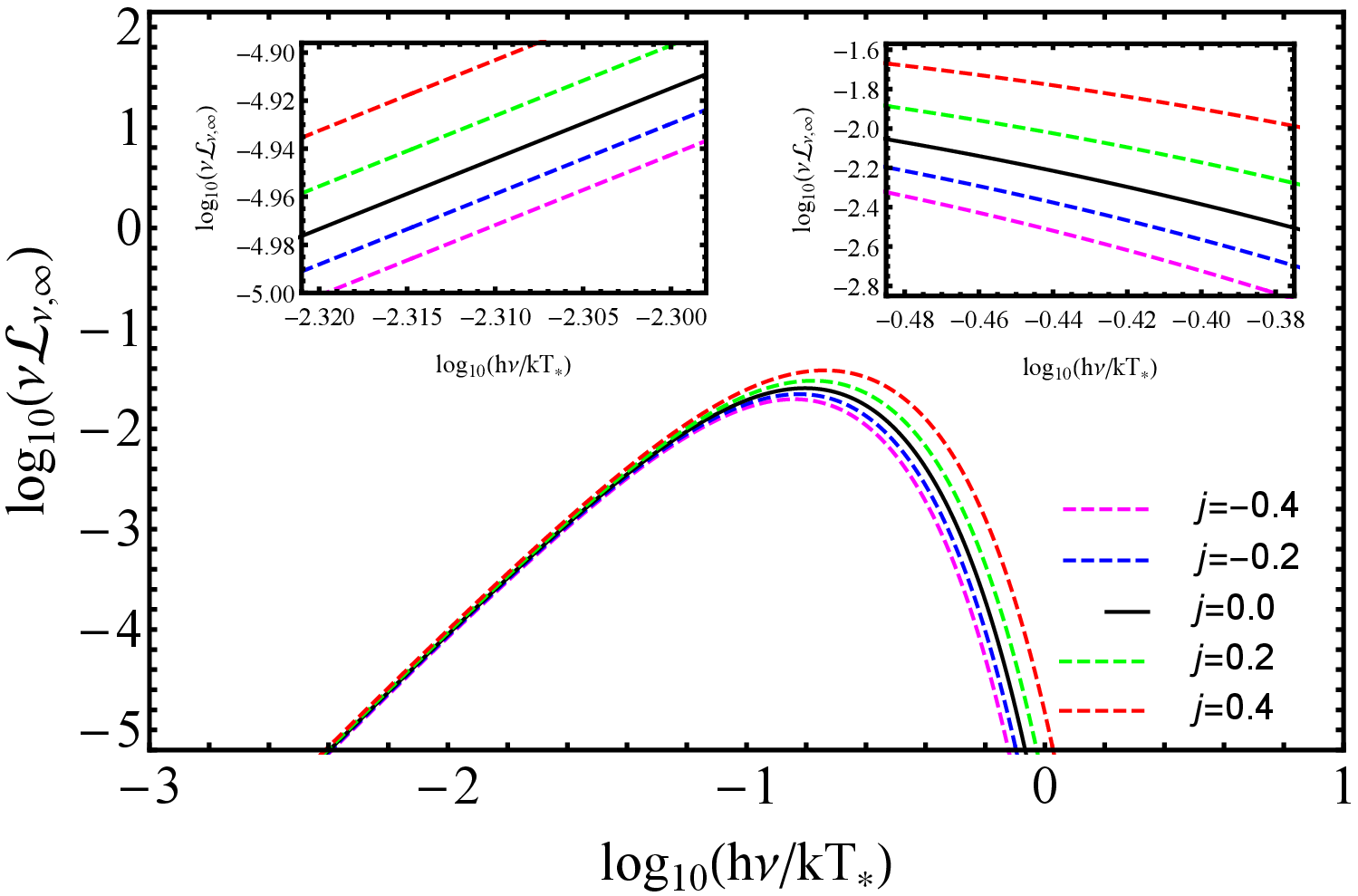}\\ } 
\end{minipage}
\hfill 
\begin{minipage}{0.50\linewidth}
\center{\includegraphics[width=0.97\linewidth]{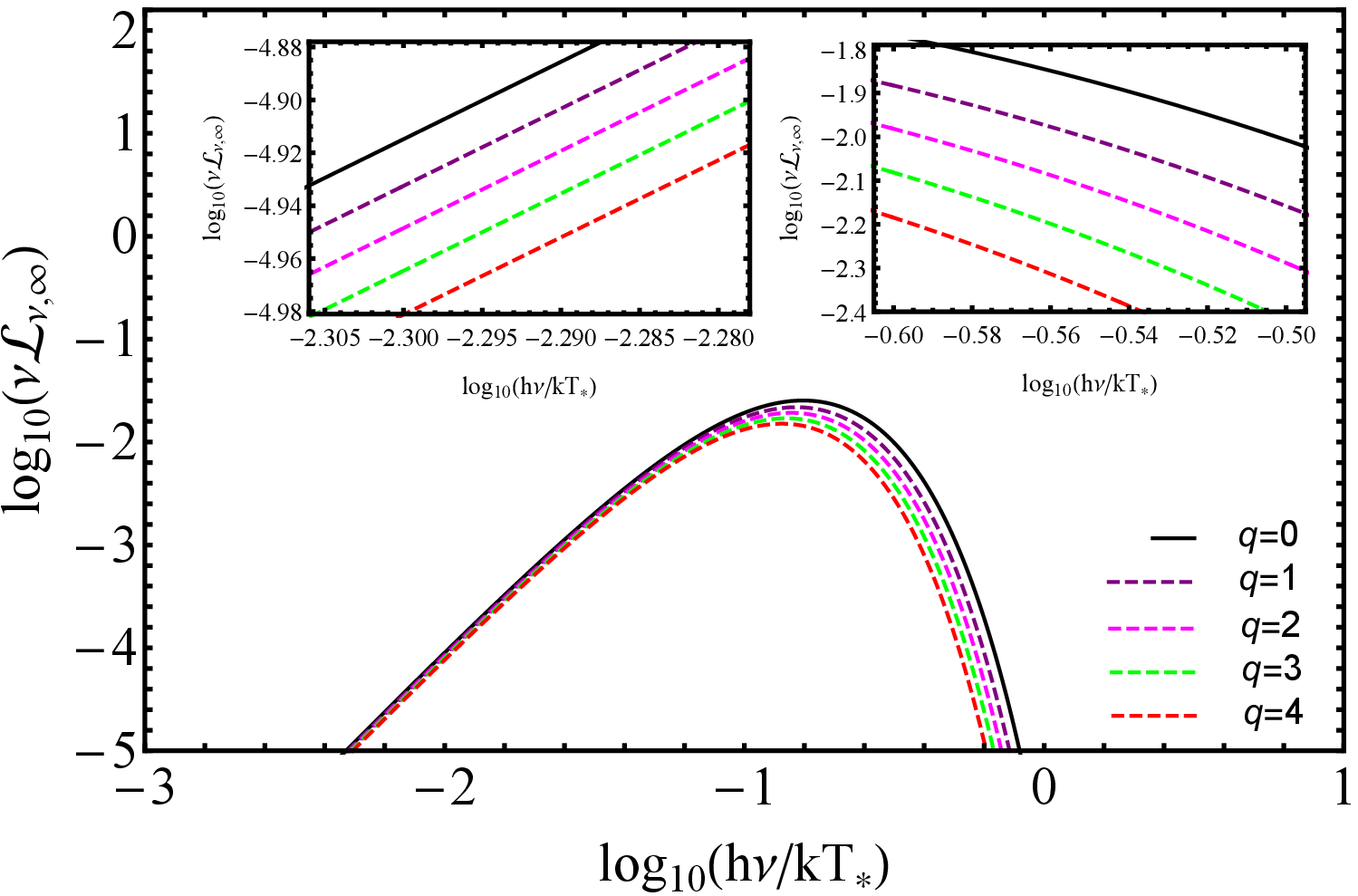}\\ }
\end{minipage}
\caption{Spectral luminosity versus frequency of the emitted radiation for blackbody emission of the accretion disk in the Hartle-Thorne metric. Left panel: Spectral luminosity versus frequency  for fixed $q=0$ and arbitrary $j$. Right panel: Spectral luminosity versus frequency  for fixed $j=0$ and arbitrary $q\geq0$.}
\label{fig:speclum}
\end{figure*}

\subsection{Numerical analysis}

We can now proceed with the numerical analysis based on the theoretical predictions derived from our metric and the assumption of a thin disk. As a result, we present the main outcomes of our study in Figures \ref{fig:flux}, \ref{fig:difflum}, and \ref{fig:speclum}. These results can also be compared with previous findings available in the literature. 

In the following, we highlight the key characteristics of our study and, specifically,
\begin{itemize}
    \item[-] Fig. \ref{fig:flux} showcases the flux distribution, which provides insights into the amount of radiation emitted by the disk as a function of the radial coordinate.
\item[-] Fig. \ref{fig:difflum} illustrates the differential luminosity, representing the energy per unit time that reaches an observer located at infinity. This quantity is crucial in understanding the overall energy output of the accretion disk.
\item[-] Fig. \ref{fig:speclum} displays the spectral luminosity distribution, denoted as $\mathcal{L}_{\nu,\infty}$. It characterizes the energy emitted by the accretion disk as a function of the frequency of radiation.

\end{itemize}

By examining these figures, we can gain valuable insights into the behavior and properties of the accretion disk, while also comparing our findings with existing literature.

In particular, for Fig. \ref{fig:flux}, we underline that

\begin{itemize}
    \item[-] the left panel corresponds to the case of fixed $q=0$ and arbitrary $j$. We observe that the flux for the Schwarzschild black hole is consistently lower (higher) than the flux for the Hartle-Thorne metric  with $j>0$ ($j<0$) in the entire range of the radial distance. The behavior of the flux in the Hartle-Thorne metric is similar to that observed in the Kerr metric for co-rotating (counter-rotating) orbits, as noted in the Ref. \cite{2021PhRvD.104h4009B};
    \item[-] the right panel shows a comparison of the flux for the Schwarzschild black hole with that of a static and deformed object characterized by $j=0$ and $q\geq0$. We find that the flux for the Schwarzschild black hole is always greater than the flux for the static and deformed object. This behavior is consistent with the behavior observed in the Kerr metric but for counter-rotating orbits, as indicated in the reference \cite{2021PhRvD.104h4009B};
    \item[-] these results provide valuable insights into the radiative flux behavior in the Hartle-Thorne metric, and they demonstrate similarities and differences with respect to other metrics, such as the Kerr metric, for various orbit configurations.
\end{itemize}

Concerning Fig. \ref{fig:difflum}, we highlight that

\begin{itemize}
    \item[-] we present the plot of the differential luminosity as a function of the normalized radial coordinate in the Hartle-Thorne metric. The left panel corresponds to the case of fixed $q=0$ and arbitrary $j$, while the right panel represents the case of fixed $j=0$ and arbitrary $q\geq0$. As the differential luminosity is directly defined in terms of the flux, we notice  that the behavior observed in Fig.~\ref{fig:flux} naturally translates into the behavior of the differential luminosity. The left panel of Fig.~\ref{fig:difflum} reflects the same trends observed in the left panel of Fig.~\ref{fig:flux}, with the differential luminosity for the Hartle-Thorne metric exceeding (not exceeding) for $j>0$ ($j<0$)  that of the Schwarzschild black hole throughout the radial distance range;
    \item[-] similarly, the right panel of Fig.~\ref{fig:difflum} reflects the behavior observed in the right panel of Fig.~\ref{fig:flux}. Here, we compare the differential luminosity for the Hartle-Thorne metric, characterized by $j=0$ and $q\geq0$, with that of a Schwarzschild metric (black solid curve). The results show that the differential luminosity for the Schwarzschild metric surpasses that of the Hartle-Thorne metric  with $j=0$ and $q\geq0$, describing gravitational fields of static and deformed objects;
    \item[-] these findings highlight the connection between the flux and the differential luminosity, confirming that the behavior observed in Fig.~\ref{fig:flux} is accurately translated into the differential luminosity as well.
\end{itemize}

While, finally, for Fig. \ref{fig:speclum}, 

\begin{itemize}
    \item[-] we present the plot of the spectral luminosity $\mathcal{L}_{\nu,\infty}$ as a function of the frequency of radiation emitted by the accretion disk in the Hartle-Thorne metric. The left panel corresponds to the case of fixed $q=0$, while the right panel represents the case of fixed $j=0$.
    \item[-] Similar to the spectral luminosity obtained using the Kerr metric for co-rotating and counter-rotating orbits, as referenced \cite{2020MNRAS.496.1115B,2021PhRvD.104h4009B}, we observe a similar pattern in the left and right panels of Fig.~\ref{fig:speclum}. This indicates that the spectral luminosity in the Hartle-Thorne metric follows a similar trend as in the Kerr metric for the respective orbit configurations.
    \item[-] Furthermore, it is noteworthy that the spectral luminosity of the accretion disk with $q=0$ and $j>0$ ($j<0$) (left panel) is always larger (smaller) than that in the Schwarzschild spacetime. Conversely, for the case of $j=0$ and $q\geq0$ (right panel), the spectral luminosity is always smaller than in the Schwarzschild spacetime.
    \item[-] These results highlight the differences in spectral luminosity between the Hartle-Thorne metric and the Schwarzschild spacetime, reaffirming the influence of the Hartle-Thorne metric parameters on the emitted radiation from the accretion disk.
\end{itemize}

\section{Neutron star models} \label{sec:ns}

In this section, we examine models of neutron stars and calculate feasible parameters for $M$, $j$, and $q$ using the Hartle-Thorne formalism. We establish that the values of $M$, $j$, and $q$ discussed above align with realistic neutron star models. Furthermore, we illustrate that the radius of ISCO, $R_{ISCO}$, can surpass the size of a neutron star.

\subsection{Mass-radius relations for neutron stars}

To determine the mass-radius relationship for neutron stars, it is necessary to select an equation of state and subsequently solve the Tolman-Oppenheimer-Volkoff equation, which describes a static neutron star in hydrostatic equilibrium. This straightforward method becomes more intricate when considering the rotation of a star. Fortunately, there exist established methodologies and publicly accessible numerical codes both in the literature and on dedicated websites, simplifying the process. These resources offer valuable tools for studying the properties and characteristics of rotating neutron stars, see e.g. \cite{2003LRR.....6....3S,2014PhRvD..89l4013Y,2015PhRvD..92b3007C}. While the technical aspects of studying neutron stars may be relatively clear, the conceptual challenges arise when considering the equation of state.

Neutron stars are extraordinary objects characterized by extreme densities, pressures, temperatures, electromagnetic fields, and gravitational fields. Unfortunately, these extreme conditions cannot be replicated or reproduced in laboratory settings. Consequently, the equation of state for neutron stars remains an area of ongoing research and is not yet firmly established\footnote{The equation of state describes the relationship between various properties of matter within a neutron star, such as density, pressure, temperature, composition etc. Understanding and accurately modeling these relationships is crucial for comprehending the internal structure and behavior of neutron stars. Despite significant progress, the equation of state for neutron stars remains an active area of research and a subject of ongoing debate and investigation and, so, determining an equation of state that accurately represents the complex physics at play is a challenging task. }.

There are a lot of uncertainties at supra-nuclear densities \cite{1971reas.book.....Z,1983bhwd.book.....S, haenselbook} and this fact generates various models~\cite{2001ApJ...550..426L, 2004Sci...304..536L, 2007PhR...442..109L,2016PhR...621..127L,2020MNRAS.495.5027P,2020ApJ...901..155G}. In this work we adopt the neutron star model formulated in Ref.~\cite{2012NuPhA.883....1B}. We consider this approach since: 

\begin{itemize}
    \item[-] the model or equation of state takes into account strong interaction (based on the Boguta-Bodmer model \cite{1977NuPhA.292..413B}), weak interaction (taking into account the $\beta$-equilibrium), electromagnetic and gravitational interactions by solving the Einstein-Maxwell-Thomas-Fermi system of equations;
    \item[-] the equation of state fulfills the global charge neutrality condition, unlike other equations of state which are mainly derived to satisfy the local charge neutrality condition;
    \item[-] the equation of state fulfills both theoretical and observational constraints
\end{itemize}

These are the main reason and motivations to exploit this model. Certainly, it should be stressed that one can refer to any model in the literature. In order to show that our findings are compatible with neutron star physics, and to estimate the basic parameters of neutron stars we adopt here the above mentioned model.

Hence, to construct the mass-radius, mass-central density, radius-central density and other relation we exploit the well-corroborated Hartle's formalism \cite{1967ApJ...150.1005H,1968ApJ...153..807H}.

\begin{figure}[ht]
\includegraphics[width=\columnwidth]{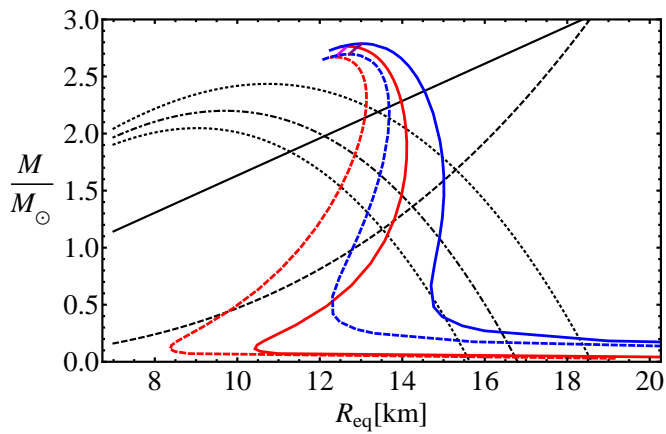}
\caption{The mass-radius relations for neutron stars. Dashed/solid blue and red curves indicate static/rotating local charge neutrality (LCN) and global charge neutrality (GCN) cases, respectively. The solid line is the upper limit of the surface gravity of XTE J1814-338, the dotted-dashed curve show the lower limit to the radius of RX J1856-3754, the dashed line is the constraint imposed by the fastest rotating pulsar PSR J1748-2246ad and the dotted curves are the 90 \% confidence level contours of constant $R_\infty$ of the neutron star in the low mass X-Ray binary X7. Any realistic mass-radius relation should pass through the area delimited by the solid, the dashed and the dotted lines and in addition it must have a maximum mass larger than the mass of PSR J0740+6620, $M=(2.14\pm0.2)M_{\odot}$. }
\label{fig:MR}
\end{figure}

Specifically, in Fig.~\ref{fig:MR}, the mass-radius relation is shown along with observational constraints. The neutron star model we adopted fully satisfies the observational and theoretical constraints, which are listed below\footnote{For additional details, see e.g.~\cite{2014NuPhA.921...33B}.}. 

In the future, it is anticipated that we may observe more massive neutron stars, as both theoretical static and rotating masses have already exceeded the maximum observed mass. These potential discoveries could provide valuable insights into the nature of neutron star equations of state, particularly those characterized by stiffness or super stiffness \cite{2016IJMPA..3141017Y}. By studying these extreme objects, we can gain a deeper understanding of the fundamental properties of matter under extreme conditions, pushing the boundaries of our knowledge in the field of astrophysics. 

\subsection{Theoretical and observational constraints}

In order to construct realistic and physically self-consistent neutron star models, certain constraints, both theoretical and observational, must be satisfied. Theoretical constraints are derived from fundamental principles and govern the maximum mass of a neutron star, as defined by the Tolman-Oppenheimer-Volkoff limit. However, this limit is model-dependent and needs to be considered in conjunction with observational constraints on neutron star masses \cite{1996A&A...305..871B,1996ApJ...470L..61K,2010Natur.467.1081D,2018ApJ...852L..25R}. Another important theoretical constraint is related to the speed of sound in the extremely dense matter within a neutron star \cite{tlemissov2020analysis}. To satisfy the causality principle, the speed of sound must not exceed the speed of light in vacuum. For rotating neutron stars with crusts, there exists a constraint on the spin parameter $j$, which should not exceed 0.7 \cite{2011ApJ...728...12L}. However, for crustless neutron stars, the spin parameter can approach or even exceed unity \cite{2016RAA....16...60Q}. In our adopted neutron star model, we encounter similar circumstances, where locally neutral neutron stars exhibit properties akin to those with crusts, while globally neutral neutron stars resemble those without crusts \cite{2018mgm..conf.3433B}.

Observational constraints on the mass-radius relations of neutron stars are derived from various measurements, including the largest observed masses \cite{2013Sci...340..448A,2020NatAs...4...72C}, the largest observed radii \cite{2004NuPhS.132..560T}, the highest rotational frequencies \cite{2004Sci...304..536L,2006Sci...311.1901H}, and the maximum surface gravity \cite{2006ApJ...644.1090H,2011PrPNP..66..674T}. These observational constraints play a crucial role in validating and refining our understanding of neutron star properties. For a visual representation, refer to Fig.~\ref{fig:MR}.

\subsection{Morphological properties of maximally rotating neutron stars}
\begin{figure}[ht]
\includegraphics[width=\columnwidth]{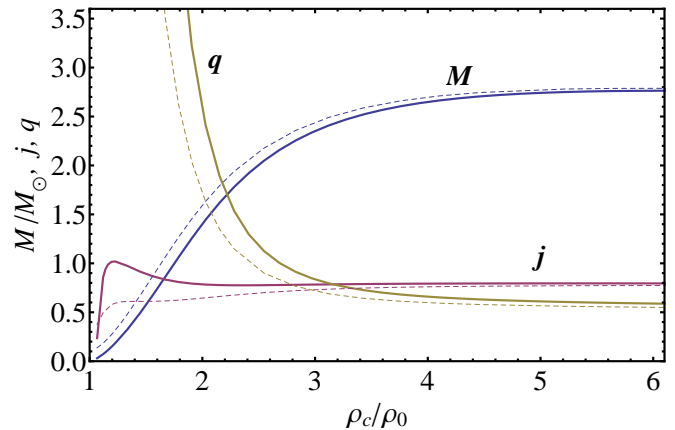}
\caption{The mass in solar masses ($M/M_{\odot}$), the dimensionless angular momentum $j$ and quadrupole moment $q$ as a function of the central density of a maximally rotating neutron star. Solid curves indicate GCN and dashed curves indicate LCN cases.}
\label{fig:lgcn}
\end{figure}

As an intriguing point, we can now work out the morphological properties of maximally rotating neutron stars, employing the main characteristics of mass, angular momentum and so forth. 

In particular, Fig. \ref{fig:lgcn} illustrates the key parameters of neutron stars as a function of the central density normalized by the nuclear density ($\rho_0\approx2.3\times 10^{17}$ kg/m$^3$) and,

\begin{itemize}
    \item[-] the mass, dimensionless angular momentum ($j$), and quadrupole moment ($q$) are depicted for maximally rotating configurations, calculated using the Hartle-Thorne formalism and employing the equation of state described in References \cite{2012NuPhA.883....1B,2015ARep...59..441B}
    \item[-] the plot reveals that as the central density increases, the mass of the neutron star reaches its maximum value, but both $j$ and $q$ decrease. For central densities beyond $\rho_c/\rho_0\approx3.4$, the angular momentum parameter $j$ becomes dominant over the quadrupole moment parameter $q$. This intriguing effect suggests that as the central density of a neutron star increases, it becomes more spherical in shape but rotates at a faster rate; 
    \item[-] on the other hand, at lower density ranges, the influence of the quadrupole moment $q$ becomes more pronounced than that of $j$. Consequently, rotating and more massive neutron stars exhibit similarities to Kerr black holes.
\end{itemize}

\begin{figure}[ht]
\includegraphics[width=\columnwidth]{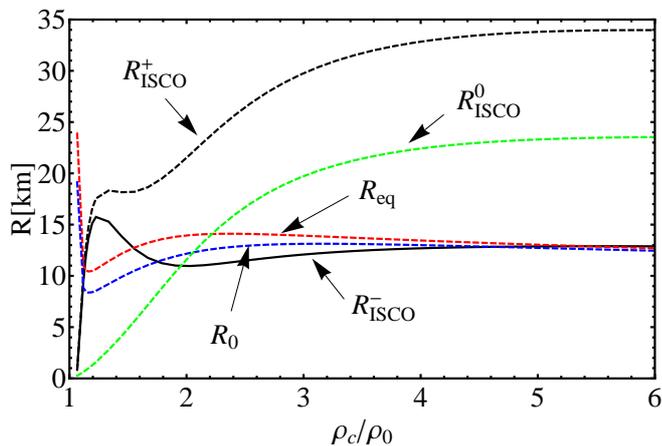}
\caption{Different radii as a function of the normalized central density for neutron stars. Where $R_{ISCO}^-/R_{ISCO}^+$ are the ISCO radii for prograde/retrograde orbits, $R_{eq}$ is the equatorial radius of maximally rotating NSs, $R_0$ is the static radius for spherical non-rotating NSs, and $R_{ISCO}^0=6M$ with $M$ being the static mass of NSs.}
\label{fig:radii}
\end{figure}

In addition, Fig. \ref{fig:radii} presents the relation between the radius and central density of neutron stars, including the $R_{ISCO}$ and,

\begin{itemize}
    \item[-] in the case of a static neutron star, the equatorial radius $R_{eq}$ corresponds to the static radius $R_0$. However, starting from a central density of approximately $\rho_c/\rho_0\approx2.2$, which corresponds to a neutron star with a mass of approximately $2M_{\odot}$, the static radius $R_0$ becomes smaller than the innermost stable circular orbit radius $R_{ISCO}^0$. This implies that when measuring $R_{ISCO}^0$, or alternatively the flux and luminosity, it becomes challenging to distinguish between massive neutron stars and low-mass static black holes. To differentiate a massive neutron star from a black hole, one possibility is to observe the photon sphere, which can provide distinctive features;
    \item[-] regarding the innermost stable circular orbit radius for co-rotating disks ($R_{ISCO}^-$), if a neutron star rotates rapidly, $R_{ISCO}^-$ tends to shrink towards the surface of the star. Only for higher central densities does it become slightly larger than the equatorial radius ($R_{eq}$) and exhibit comparable characteristics to a rotating black hole;
    \item[-]  on the other hand, for counter-rotating disks, the innermost stable circular orbit radius ($R_{ISCO}^+$) is larger than $R_{eq}$ and $R_{ISCO}^0$. It is worth noting that, from a theoretical point of view, counter-rotating disks may exist. However, they are less likely and viable due to the frame-dragging effect near relativistic objects such as neutron stars and black holes. The presence of counter-rotating  accretion disks around compact objects remains a topic of ongoing research and investigation.
\end{itemize}

Finally, it is worth noting that in Figs.~ \ref{fig:radii}, \ref{fig:lgcn} and \ref{fig:MR} we considered maximally rotating neutron stars, but most of the observed neutron stars and pulsars do not rotate so fast as theoretically estimated \cite{2014NuPhA.921...33B}. Therefore, for slowly rotating stars $j$, $q$ are quite small, and $R_{ISCO}^-$ can be close to $R_{ISCO}^0$ i.e. outside a neutron star. This implies only one thing that the values of $j$ and $q$ we used above are sufficiently realistic.


\section{Final outlooks and perspectives}\label{sez6}

In the present work, we investigated the motion of neutral test particles in the Hartle-Thorne spacetime by considering equatorial circular geodesics and examining the influence of the central object angular momentum and quadrupole moment.

We computed important orbital parameters, including angular velocity, angular momentum, and energy for neutral test particles on circular orbits. These quantities exhibited smaller (larger) values for $q=0$ and $j>0$ ($j<0$), ($j=0$ and $q>0$) compared with the Schwarzschild case. The ISCO radius was also estimated, and we calculated the efficiency of matter-to-radiation conversion for different values of $j$ and $q$. Our results showed that highly rotating and less deformed objects were more efficient in converting matter into radiation than the Schwarzschild black hole of the same mass, and vice versa.

Using the Novikov-Page-Thorne thin accretion disk model, we determined the radiative flux, differential luminosity, and spectral luminosity. Contrary to angular velocity, angular momentum, and energy, these quantities exhibited larger (smaller) values for $q=0$ and $j>0$, ($j<0$) ($j=0$ and $q>0$) compared with the Schwarzschild case.

Afterwards, we employed a neutron star model described in Refs.~\cite{2012NuPhA.883....1B, 2014NuPhA.921...33B} to calculate various key parameters of rotating neutron stars, including mass-radius relations, mass-central density relations, radius-central density relations, angular momentum, and quadrupole moment. We compared the ISCO radius with realistic neutron star radii and found that for slowly rotating massive neutron stars, the ISCO radius lies outside the star. Furthermore, we demonstrated that the values of $j$ and $q$ utilized in our study are realistic and consistent with neutron star physics.

Our findings exhibited strong compatibility with configurations incorporating quadrupole effects, obviating the need for additional rotational parameters. However, our results aligned more closely with the Kerr spacetime compared to other metrics. Notably, the $q$-metric failed to reproduce comparable outcomes to the Hartle-Thorne spacetime. 

In view of these remarkable results, we intend to further investigate this metric and explore other properties that display notable distinctions from metrics employing quadrupole effects. In particular, involving more accurate accretion disk models and the Hartle-Thorne spacetime would help in shedding light on the role played by quadrupole moments in compact object physics.

\begin{acknowledgments}

YeK acknowledges Grant No. AP19575366, TK acknowledges Grant No. AP19174979  and KB, OL, and MM acknowledge Grant No. AP19680128 from the Science Committee of the Ministry of Science and Higher Education of the Republic of Kazakhstan. 
KB is grateful to the Departments of Physics and Mathematics at the University of Camerino for academic mobility provided by Erasmus+ program ``I CAMERIN01'' (2022-1-IT02-KA171-HED-000073309), during the period in which this manuscript has been written. He is particularly grateful to prof. Carlo Lucheroni for his economical support during the period in which the paper has been written.
OL is grateful to INAF, National Institute of Astrophysics, for the support and in particular to Roberto della Ceca, Gaetano Telesio and Filippo M. Zerbi for discussions. It is also a pleasure to acknowledge Carlo Cafaro and Roberto Giamb\`o for fruitful discussions on the subject of this paper. The work of HQ was partially supported  by UNAM-DGAPA-PAPIIT, Grant No. 114520, and CONACYT-Mexico, Grant No. A1-S-31269.
\end{acknowledgments}

\end{document}